\documentclass[fleqn,usenatbib]{mnras}

\usepackage[T1]{fontenc}
\usepackage{ae,aecompl}

\usepackage{graphicx}	
\usepackage{amsmath}	
\usepackage{amssymb}  
\usepackage{subcaption}
\usepackage{xcolor}
\usepackage{float}
\restylefloat{table}
\usepackage{ulem}

\newcommand\lsim{\mathrel{\rlap{\lower4pt\hbox{\hskip1pt$\sim$}}
        \raise1pt\hbox{$<$}}}
\newcommand\gsim{\mathrel{\rlap{\lower4pt\hbox{\hskip1pt$\sim$}}
        \raise1pt\hbox{$>$}}}
\newcommand{\cmfast}{\textsc{\small 21cmFAST}}
\newcommand{\cmsense}{\textsc{\small 21cmSENSE}}
\newcommand{\delT}{\delta T_{\rm b}}

\title[Minimum size of 21-cm simulations]{Minimum size of 21-cm simulations}
\author[Kaur et. al.]{
Harman Deep Kaur\thanks{E-mail: harman.kaur@sns.it},
Nicolas Gillet,
Andrei Mesinger\\
$^{1}$Scuola Normale Superiore, Piazza dei Cavalieri, 7 56126 Pisa, Italy
}

\date{Accepted XXX. Received YYY; in original form ZZZ}

\pubyear{2019}

\begin{document}
\label{firstpage}
\pagerange{\pageref{firstpage}--\pageref{lastpage}}
\maketitle

\begin{abstract}
  
  Cosmic 21cm interferometry is set to revolutionize our understanding of the Epoch of Reionization (EoR) and the Cosmic Dawn (CD).  However, the signal has structure on a huge range of scales, requiring large simulation boxes to statistically capture the relevant fields. In this work we quantify the minimum box size for simulating the power spectrum (PS) of the cosmic 21cm signal.  We perform multiple realizations of the initial conditions, for a range of box sizes.  We quantify convergence with respect to a mock observation of box length $1.1$ Gpc, with thermal noise computed for a 1000h observation with SKA1-low. We find that simulations of box lengths $ L \sim 200$--300 Mpc underestimate the large-scale power during the CD by $\sim$ 7--9 \% on average. 
We conclude that box lengths of $L \gsim$ 250 Mpc are needed to converge at the level of $\lsim$ 1 $\sigma$ of the total noise.  
\end{abstract}




\section{Introduction}
\label{sec:intro}

The cosmic 21-cm signal will be a powerful probe of the Cosmic Dawn (CD) and subsequent Epoch of Reionization (EoR; see e.g. \citealt{Furlanetto2006, Pritchard2012, Mesinger2019}).
The signal is commonly expressed as the difference of the brightness temperature of cosmic gas ($\delta T_{\rm b}$) with respect to the radio background, commonly taken to be the CMB temperature ($T_{\gamma}$) :
\begin{align}
      \label{eq:delT}
      \nonumber \delta T_{\rm b} &\approx 27x_{\rm HI}(1+\delta)\left(\frac{H}{dv_{r}/dr+H}   \right)\left(1- \frac{T_{\gamma}}{T_{\rm S}}\right)\\
    &\times \left(\frac{1+z}{10} \frac{0.15}{\Omega_{\rm M}\rm h^2}\right)^{1/2}\left(\frac{\Omega_{\rm b}\rm h^2}{0.023}\right) \ \rm mK .
\end{align}
here $x_{\rm HI}$ is the fraction of neutral hydrogen, $\delta$ is the gas overdensity ($\delta = \rho/\overline{\rho}-1$), $H$ is the Hubble parameter, $dv_{r}/dr$  is the peculiar velocity gradient along the line-of-sight, $T_{\rm S}$ is the spin temperature defined by the relative abundances of the excited and ground states of the spin-flip transition. 

We can see from eq. (\ref{eq:delT}) that the signal is sensitive to cosmology as well as the ionization and thermal state of the intergalactic medium (IGM).  During the CD and EoR, these gas properties are governed by radiation fields from rare, nascent galaxies.  UV ionizing photons from stars and black holes inside the first galaxies ultimately reionize the Universe (e.g. see the review in \citealt{Mesinger16} and references therein).
Before the EoR, X-rays from high mass X-ray binaries or the hot ISM likely dominated the heating of the IGM (e.g. \citealt{Furlanetto06, Baek10, Santos10, McQuinn12, Pacucci14, MFS17, Ross17, Eide18}).  Moreover, the Lyman alpha background is responsible for coupling the spin and kinetic temperatures of the IGM during the CD \citep{Wouthuysen52, Field59}.  Therefore we can use upcoming 21-cm observations to infer the ionizing, X-ray and soft UV properties of the first galaxies (e.g. \citealt{GM17_21CMMC}).

These first galaxies would be hosted by rare and highly biased dark matter halos, whose abundances are modulated by long-wavelength modes of the density field (e.g. the so-called ``peak-patch'' formalism; \citealt{BM96_algo}).  Thus the number density of galaxies, and correspondingly the emissivity of radiation, can fluctuate dramatically on scales of tens of Mpc in the early Universe.

This can have profound implications on the EoR and CD.  Because ionizing photons have a short mean free path and the typical recombination times in the IGM are long, the patchiness of the EoR can be directly related to the patchiness of the galaxy fields (e.g. \citealt{FZH04}).  Numerically simulating galaxy fields for highly biased sources requires large-scale boxes, in order to capture the relevant long-wavelength density modes. 
Using analytic, conditional halo mass functions, \citet{Barkana2004} showed that small-box simulations result in an EoR which occurs too rapidly, too homogeneously, and too late.
\citet{Iliev2014} further quantified this bias during the EoR using numerical radiative transfer (RT) simulations, estimating that boxes of $\gsim200$ Mpc would be required for convergence in the EoR and corresponding 21-cm power spectrum.

\begin{figure}
    \centering
    \includegraphics[width=\columnwidth]{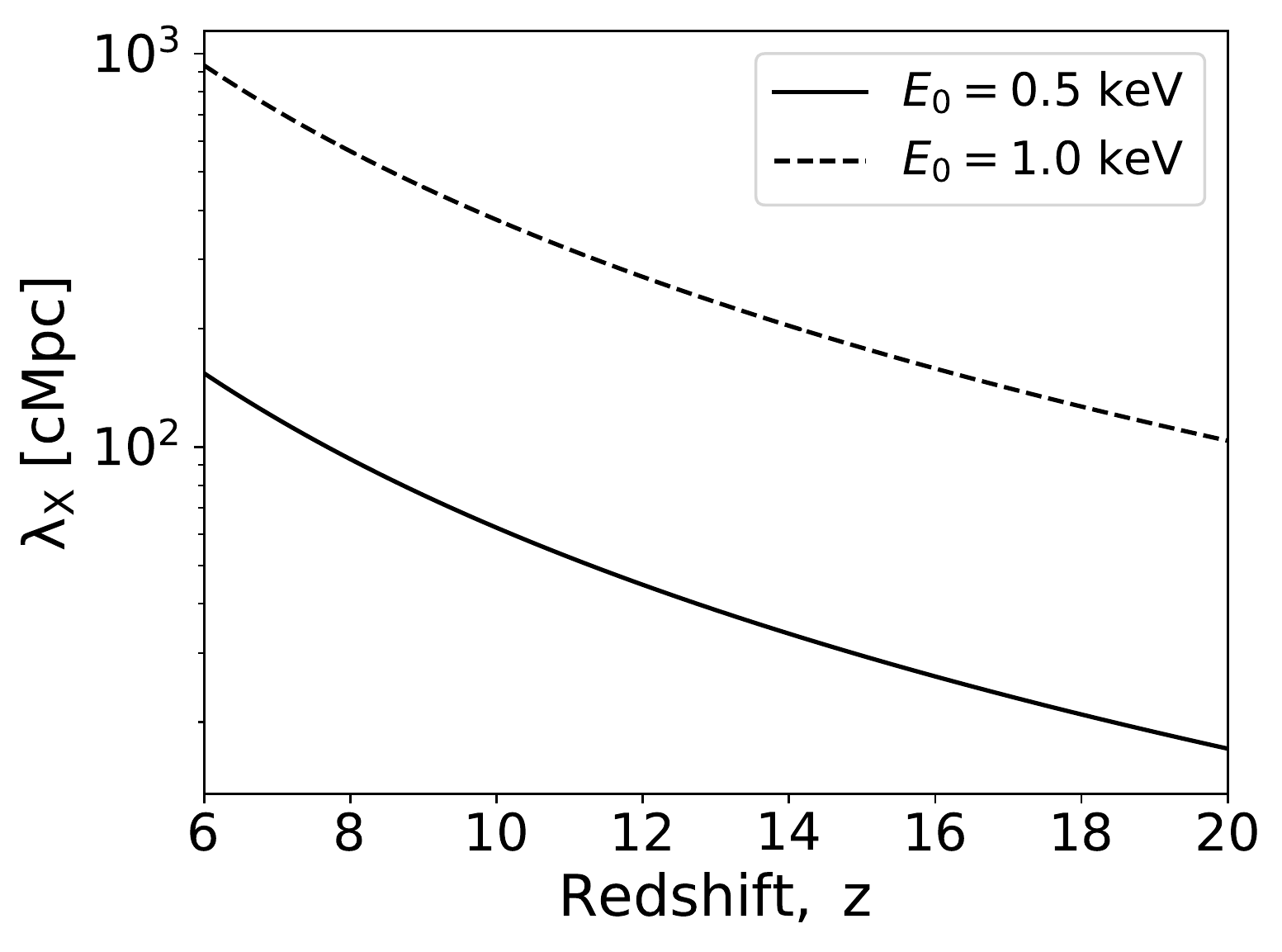}.
    \caption{The mean free path of X-ray photons through a mean density, neutral IGM.  Curves correspond to photon energies of 0.5 keV and 1 keV (c.f. eq. \ref{eq:mfp}).
    \label{fig:MFP}}
\end{figure}

But what about the earlier stages of the CD?  These epochs are driven by soft UV and X-ray photons from even more biased galaxies.  Unlike ionizing photons, these long mean free path photons are capable of interacting with the IGM over a wide range of scales.
For example, the mean free path of X-rays in the high-$z$ IGM (e.g. \citealt{Mcquinn2012}),
\begin{equation}
\label{eq:mfp}
  \lambda_{\rm X} \approx 20{\overline{x}_{\rm HI}}^{-1} {\left[\frac{E_{\rm X}}{300\rm eV}\right]}^{2.6}{\left[\frac{1+z}{10}\right]}^{-2} \rm cMpc ,
\end{equation}
is a strong function of the photon energy $E_X$ (see Fig. \ref{fig:MFP}).
The corresponding large range of relevant scales, modulated by the highly biased first galaxies, results in large-scale ($k\lsim$0.1 Mpc$^{-1}$) fluctuations in the 21-cm power spectrum during the CD (e.g. \citealt{Pritchard2007}).  This suggests that we might need even larger simulation boxes to model the CD.  Because multi-frequency RT simulations are very computationally expensive, it is important to know what is the smallest box size capable of accurately simulating the signal.

{\it Here we quantify how large does a simulation need to be to capture the cosmic 21-cm signal.}  Using the public simulation code \cmfast\ \citep{Mesinger2007, Mesinger2011}, we perform a convergence study of the 21-cm power spectrum (PS) summary statistic.  Varying the initial seeds of our simulations, we show how decreasing the box size increases scatter in the 21-cm PS, for the same astrophysical model (e.g. \citealt{Mondal16}).  We quantify convergence using a mock 1000h observation from the upcoming Square Kilometre Array (SKA) telescope, phase 1.  We find that box sizes of $\gsim 250$ Mpc are needed for the simulated PS to converge to within $\lsim 1 \sigma$ of the total noise.

The outline of the paper is as follows. We begin by describing the methodology of our simulations, the astrophysical model, and the calculation of noise, in \S \ref{sec:method}. 
In \S \ref{sec:results}, we show the results of our convergence studies, before concluding in \S \ref{sec:conc}.
Throughout the paper, we assume a $\Lambda$CDM cosmology with ($\Omega_{\rm M} = 0.308,\ \Omega_{\rm b} = 0.0484, \ \Omega_{\Lambda}=0.692,\ \sigma_{8} = 0.815, \ \rm h = 0.678, \ \rm n_{\rm s} = 0.968  $), consistent with results from the {\it Planck} telescope \citep{Planck2016}.

\section{Methodology}
\label{sec:method}

We simulate the 21-cm signal during the CD and EoR using the public code \cmfast\ v2.1 \citep{Mesinger2007, Mesinger2011}\footnote{\url{https://github.com/andreimesinger/21cmFAST}}, and create mock telescope noise using the public package \cmsense\ \citep{Pober2013b, Pober2014}\footnote{\url{https://github.com/jpober/21cmSense}}.  Our reference simulation and mock observation are generated using a box of $\sim$ 1.1 Gpc per side.  We then run suites of smaller box simulations and quantify convergence with respect to the reference.  Here we briefly summarize the methodology, encouraging readers to consult the afore-mentioned references for more details.

\subsection{Modelling the 21-cm signal}

\subsubsection{IGM properties}

\cmfast\ samples a cosmological PS to create a realization of a Gaussian random field, and then evolves this realization with second order perturbation theory (e.g. \citealt{Scoccimarro1998}) in order to create density and velocity fields at the desired redshifts.

Due to the short mean free path of ionizing photons in the neutral IGM and typically-long recombination time-scales, reionization proceeds in a bi-modal fashion (e.g. \citealt{TG11} and references therein): (nearly) fully-ionized cosmic HII regions appear around nascent galaxies and expand into the (nearly) fully neutral IGM.

Cosmic HII regions are identified following the excursion set procedure \citep{FZH04}.  We compare the cumulative number of ionizing photons per baryon, $\bar{n}_{\rm ion}$, to the number of recombinations per baryon, $\bar{n}_{\rm rec}$, averaged over spheres of decreasing radii around a gas element.  Thus a cell at a spatial position ${\bf x}$ and redshift $z$ is marked as ionized if:
\begin{equation}
  \label{eq:nion}
    \bar{n}_{\rm ion}(\mathbf{x} , z) \geq (1+\overline{n}_{\rm rec})(1-\overline{x}_{\rm e}),
\end{equation}
where the final term accounts for pre-ionization by X-rays (discussed below).  Inside the ionized IGM, inhomogeneous recombinations and the residual HI fraction is computed according to \citet{Sobacchi2014}: assuming a temperature of $10^4$ K, a subgrid density distribution following \citet{MHR00} but adjusted for the cell's average density, and using the self-shielding prescription from \citet{Rahmati2013}.

The neutral IGM outside of the cosmic HII regions is impacted by more diffuse radiation, notably X-rays.  We follow the ionization fraction, $x_e$, and temperature evolution, $T_{\rm K}$, of the neutral IGM according to:
\begin{equation}
\frac{\rm d x_{\rm e}(\mathbf{x},z')}{\rm d z'} = \frac{\textup{d} t}{\rm d z'} \left[\Gamma_{\rm X} - \alpha_{\rm A}C x_{\rm e}^{2}n_{\rm b}f_{\rm H} \right]
\end{equation}

and 
\begin{equation}
    \frac{\textup{d} T_{\rm K}(\mathbf{x},z')}{\textup{d} z'} = \frac{2}{3k_{\rm b}(1+x_{\rm e})}\frac{\textup{d} t}{\textup{d} z'}\sum Q_{\rm p} +
    \frac{2T_{\rm K}}{3n_{\rm b}} \frac{ \textup{d} n_{\rm b}}{\textup{d} z'} - \frac{T_{\rm K}}{1+x_{\rm e}} \frac{ \textup{d} x_{\rm e}}{\textup{d} z'}
\end{equation},
where $n_{\rm b}$ is the baryon number density at $(\mathbf{x},z')$, $ \Gamma_{\rm X}$ is the ionization rate per baryon from X-rays, $\alpha_{\rm A}$ is the case-A recombination coefficient, $C$ is the sub-grid clumping factor, $f_{\rm H} $ is the hydrogen number fraction, $k_{\rm b}$ is the Boltzmann constant, and the heating rate per baryon, $Q_{\rm p}$, includes both Compton heating and X-ray heating.

The X-ray heating and ionization rates can be expressed as:
\begin{equation}
    Q_{X}(\bold{x},z) = \int d\nu \dfrac{4\pi J}{h\nu}\sum_{i}(h\nu - E_{i}^{th})f_{\rm heat}f_{i}x_{i}\sigma_{i}
\end{equation}
\begin{equation}
    \Gamma_{\rm ion,X}(\bold{x},z) = \int d\nu \dfrac{4\pi J}{h\nu}\sum_{i} f_{i} x_{i} \sigma_{i} F_{i} 
\end{equation}
 where 
\begin{align}
      F_{i} = (h\nu - E_{i}^{th})\left(\frac{f_{\rm ion,HI}}{E_{\rm HI}^{\rm th}} + \frac{f_{\rm ion,HeI}}{E_{\rm HeI}^{\rm th}} + \frac{f_{\rm ion,HII}}{E_{\rm HII}^{\rm th}}  \right) + 1
\end{align} 
 Here i stands for the atomic species: H, HeI and HII, $E_i^{\rm th}$ is their corresponding ionization threshold, $f_{i}$ their number fraction, $x_{i}$ the ionization fraction, $\sigma_i$ the cross-section, $f_{\rm heat}$ is the fraction of the primary ionized electron's energy dissipating as heat and $f_{\rm ion,j}$ is its energy contributing to secondary ionization of the species j, taken from \citet{FS10}.  The angle-averaged specific X-ray intensity, $J$, is computed by integrating the specific comoving emissivity, $\epsilon_{\rm x}$, back along the lightcone:
 \begin{equation}
   \label{eq:Jx}
    J(\mathbf{x},E,z) = \frac{(1+z)^3}{4\pi^2}\int_{z}^{\infty}dz'\frac{\rm c\it dt}{dz'}\epsilon_{\rm x} e^{-\tau},
    \end{equation}
with $e^{-\tau}$ accounting for attenuation from HI, HeI, and HeII according to \citet{Mesinger2011}.
 
Then the spin temperature $T_{\rm S}$ can be calculated as:
\begin{equation}
    T_{\rm S}^{-1}  = \frac{T_{\gamma}^{-1} + x_{\alpha}T_{\alpha}^{-1} + x_{\rm c}T_{\rm K}^{-1}}{1+x_{\alpha}+x_{\rm c}},   
\end{equation}
where $T_{\alpha} \sim T_{\rm K}$ is the color temperature set by Lyman-alpha scatterings (e.g. \citealt{Hirata06}), $x_{\alpha}$ is the WF coupling coefficient \citep{Wouthuysen52, Field59} and $x_{\rm c}$ is the collisional coupling coefficient. The Lyman alpha background used in computing $T_\alpha$ and $x_\alpha$ is computed analogously to eq. (\ref{eq:Jx}), integrating over a Pop II stellar spectrum and accounting for ``picket-fence'' absorption in the Lyman transitions.  For more details, refer to \citet{Mesinger2011}.

\subsubsection{Galaxy properties}

The source emissivities used in the previous section (e.g. equations \ref{eq:nion} and \ref{eq:Jx}) are computed according to the source model in \citet{Park2019}, which uses power-law scaling relations to relate the star formation rates (SFRs) and ionizing escape fractions to the host halo mass.
i
Specifically, the typical SFR of a galaxy in a halo of mass $M_h$ is:
\begin{equation}
{\rm SFR}(M_h, z) =  \frac{f_{\ast, 10}}{t_\ast H^{-1}} \left(\frac{M_{\rm h}}{10^{10}\rm M_{\odot}}\right)^{\alpha_\ast} \left( \frac{\Omega_{\rm b}}{\Omega_{\rm m}} \right) M_{\rm h} ~.
\end{equation}
Here, $f_{*,10}$ is the fraction of galactic gas in stars normalized to the value in halos of mass $10^{10}\ \rm M_{\odot}$, $\alpha_*$ is the corresponding power-law scaling of the stellar fraction with halo mass, $t_\ast$ is a dimensionless time-scale parameter and $H^{-1}$ is the Hubble time.  Analogously we allow the ionizing escape fraction, $f_{\rm esc}$ to be a power law function of the halo mass, with normalization $f_{\rm esc, 10}$ and power law index $\alpha_{\rm esc}$.

Then the (local) total source emissivity is computed by integrating over the conditional halo mass function \citep{ST99, Jenkins01, BL04, Mesinger2011}.
Inside this integral,  we include a halo occupation fraction quantified by a parameter $M_{\rm turn}$ in such a way that only $\exp\left[-M_{\rm turn}/M_{\rm h}\right]$ of the number of halos are hosting star-forming galaxies.  This accounts for the fact that small mass halos stop hosting galaxies due to inefficient gas accretion and/or feedback.

We assume the X-ray spectral energy distribution (SED) follows a power law,  $L_{\rm X} \propto E_x^{-1}$, and a low energy cut-off of $E_0\sim 0.5$ keV set by the typical opacity of the ISM of high redshift galaxies (e.g. \citealt{Fragos2013, Das2017}).  This X-ray emission is expected to come from either HMXBs or (less likely) the hot ISM.  As both of these sources scale with the SFR of the galaxy, we normalize our X-ray SED by the soft-band ($E_0 < E_{\rm X} < 2$ keV) X-ray luminosity per star formation rate, $L_{\rm X<2 \ keV}/{\rm SFR}$.

Our fiducial astrophysical parameters are taken from \citet{Park2019}:
$f_{*,10}$ = 0.05, \
$\alpha_*$ = 0.5,  \
$f_{\rm esc,10}$ = 0.1, \
$\alpha_{\rm esc}$ = -0.5, \
$M_{\rm turn}$ = $5*10^{8} \rm M_{\odot} $, \
$t_*$ = 0.5, \
$L_{X<2 \rm keV}/\rm SFR$ = $10^{40.5} \rm erg \ s^{-1} M_{\odot}^{-1} \ yr $,  and    \
$E_0$ = 0.5 keV.
The values of the UV parameters are consistent with the observed UV luminosity functions (e.g. \citealt{Bouvens2015a, Bouvens2016, Livermore2016, Ishigaki2017, Atek2018}) and reionization constraints from the CMB \citep{Planck2016} and high-$z$ QSOs \citep{MMO15}, while the X-ray properties are consistent with the observations of local, star-forming galaxies (e.g. \citealt{Fragos2013, MGS12_HMXB}).   In Appendix A, we also show results for a different galaxy model, finding it consistent with our fiducial results.

\subsection{Modelling the telescope noise}

We use \cmsense\ \citep{Pober2013b, Pober2014} to compute the associated thermal and cosmic variance noise from the PS of our large-scale reference model.  We assume 6h per night synthesis for a total of 1000h using the SKA1-low\footnote{\url{https://astronomers.skatelescope.org/}} (e.g. \citealt{Mellema2013, Koopmans2015}).  We expect roughly similar trends for the upcoming Hydrogen Epoch of Reionization Arrays (HERA\footnote{\url{http://reionization.org/}}; \citealt{DeBoer17}), as HERA is optimized for PS measurements and constraints using just the PS are comparable for the two instruments (e.g. compare \citealt{Park2019} and \citealt{Park2020}).  Our choice of SKA1-low is motivated by the fact that it has lower thermal noise at high redshifts, thus providing the most stringent convergence criteria, when convergence is defined in terms of the total noise.\footnote{With the eventual increase of collecting area expected in phase 2 (SKA2-low), the thermal noise component of the PS could be reduced by an additional factor of $\sim$few - 10 (e.g. \citealt{Koopmans2015}).  This might require even larger boxes than we suggest here, in order to reach the same level of accuracy measured in terms of the total noise.  However, we do not expect our conclusions to change significantly due to the fact that our convergence criteria are mostly driven by large-scale modes, for which the thermal noise component is less relevant compared to the cosmic variance (see for example Fig. 5 and the S/N panel of Fig. 2).
    }

  \cmsense\ computes the noise power spectrum for a given $k$ mode according to (e.g. \citealt{Morales2005, McQuinn06}):
  \begin{equation}
    \label{eq:thermalnoisepermode}
    P_{\rm N}(k, z)\approx X^{2}Y\frac{k^{3}}{2 \pi^{2}}\frac{\Omega '}{2t}T^{2}_{\rm sys} ~,
\end{equation}
where $X$ and $Y$ are conversion factors for bandwidths and solid angles to comoving distance, $\Omega'$ is a beam-dependent factor (e.g. \citealt{Parsons2014}),
$t$ is the integration time corresponding to the $k$-mode and  $T_{\rm sys} = 1.1 T_\mathrm{sky} + 40$mK is the system temperature of the instrument, as outlined in the SKA System Baseline Design.\footnote{\url{https://www.skatelescope.org/wp-content/uploads/2012/07/SKA-TEL-SKO-DD-001-1\_BaselineDesign1.pdf}} We use the configuration from the SKA1-low baseline design with a compact antennae core that has a maximal baseline of 1.7 km (longer and more sparsely sampled baselines are mainly for calibration purposes and add little sensitivity to the EoR signal).

Then the total noise including thermal noise and sample variance can be written as:
\begin{equation}
  \label{eq:noise}
    \delta P_{\rm N+S}(k, z)=\left(\sum_{i}\frac{1}{(P_{\rm N,\it i} + P_{21})^{2}}\right)^{-\frac{1}{2}},
\end{equation}
where the cosmological power spectrum is $P_{21}(k, z) \equiv \bar{\delT}(z)^2 \Delta^2_{21}(k, z) =  k^3/(2\pi^2 V) ~ \langle|\delta_{\rm 21}({\bf k}, z)|^2\rangle_k$, with $\delta_{21}({\bf x}, z) \equiv \delT({\bf x}, z)/ \bar{\delT}(z) - 1$, and the averaging is performed over modes $i$.  Note that this assumes the sample variance error is Gaussian distributed, which is a reasonable approximation in the modest S/N regime relevant for most observations \citep{Mondal2015}.

We use the ''optimistic model'' for foreground removal of \citet{Pober2014}. In this model, modes which are larger than the full width half max of the primary beam are assumed to be foreground dominated, and are not used when computing the power spectrum.  This fairly optimistic choice would correspondingly translate to more stringent lower limits on the allowed box size, for a given target signal-to-noise.

\section{Results}
\label{sec:results}

\subsection{Reference simulation}
\label{sec:ref}

\begin{figure*}
\includegraphics[width=0.46\textwidth, scale = 0.6]{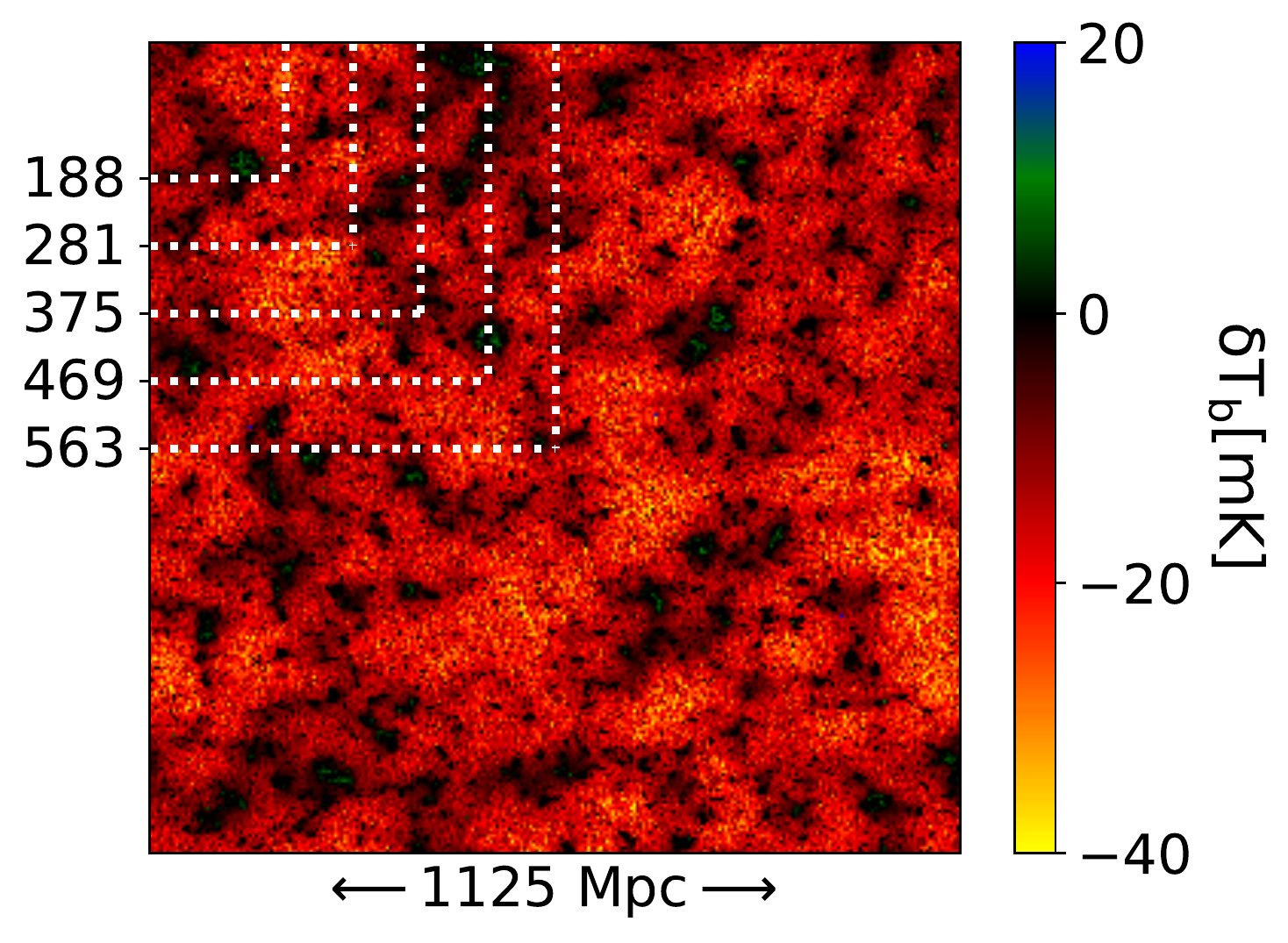}
\includegraphics[width=0.46\textwidth]{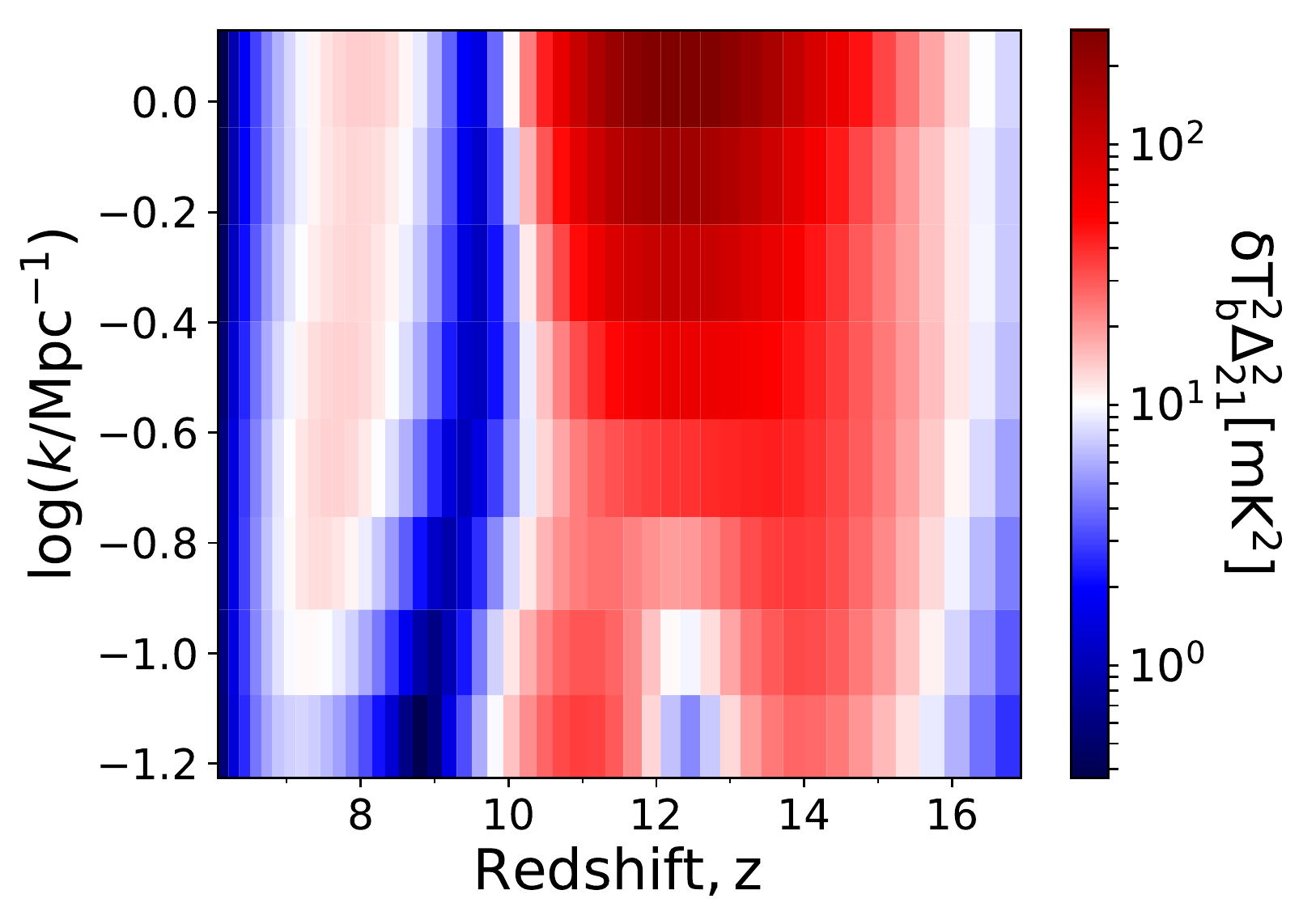}
\includegraphics[width=0.46\textwidth]{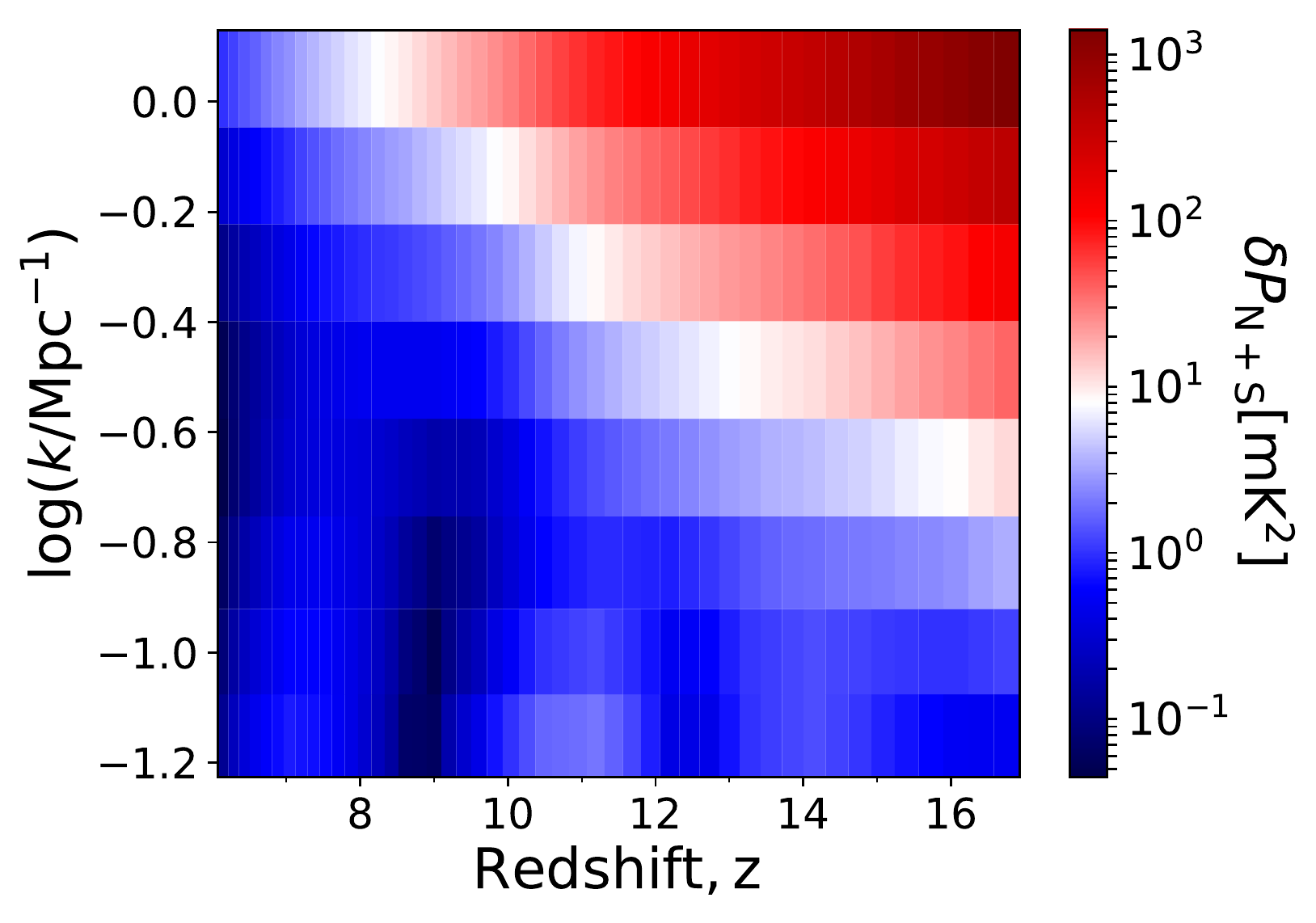}
\includegraphics[width=0.46\textwidth]{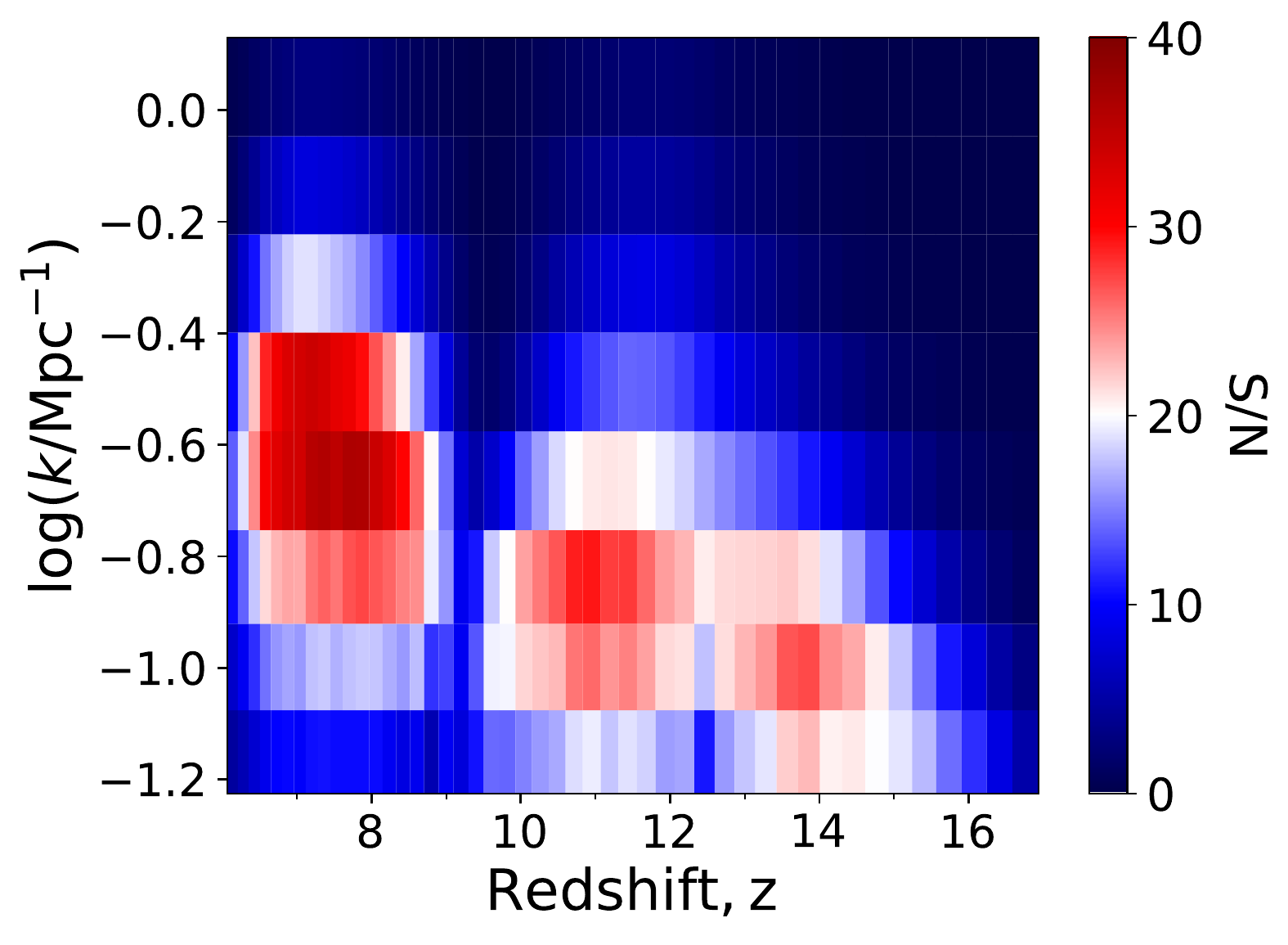}
\caption{
  {\it Top left panel}: slice through the brightness temperature at $z=10$ in our largest ``reference'' simulation, having a side length of 1125 Mpc.
  For visualization, the sizes of the smaller simulations used in this study are illustrated with white dotted lines.
  {\it Top right panel}: corresponding evolution of the 3D averaged power spectrum of the cosmic signal.
  {\it Bottom left panel}: corresponding evolution of the noise PS (including thermal and sample variance), assuming a 1000h observation with SKA1-low.
    {\it Bottom right panel}: Signal to noise (S/N), obtained by dividing the top right and bottom left panels.
  }
\label{fig:ref}
\end{figure*}

In Figure \ref{fig:ref}, we show the results from our ``reference'', large-scale simulation.  This simulation is 1125 Mpc on a side, computed on a 768$^3$ grid, and used the fiducial cosmological and astrophysical parameters discussed in the previous section.

In the top left panel we show a slice through the brightness temperature box at $z = 10$, corresponding to the late stages of the epoch of heating (EoH) for these astrophysical parameters.  Large scale fluctuations in the temperature are clearly visible from this snapshot.

In the top right panel of this figure we show the redshift evolution of the 3D averaged power spectrum.  We recover the same generic trends from previous works (e.g. \citealt{Pritchard2007, Santos10, Mesinger2011}).  Namely, the large-scale power ($k \lsim 0.1$ Mpc$^{-1}$) has three peaks in the redshift evolution ($z\sim$ 14, 11, 7 for these astrophysical parameters).  These correspond to the epochs of WF coupling, EoH, and EoR, when the large scale PS is driven by spatial fluctuations in the WF coupling coefficient, kinetic temperature, and ionization fraction, respectively.  The WF coupling and EoH peaks of the PS evolution merge on small scales, due to the stronger negative contribution of the cross-power (e.g. \citealt{Pritchard2007, MFS13}).

In the bottom left panel of Fig. \ref{fig:ref} we show the redshift evolution of the noise power spectrum, including both thermal and sample variance terms (c.f. eq. \ref{eq:noise}).  
As discussed in the previous section, the thermal component was calculated for a 1000h integration with SKA1-low, assuming optimistic foregrounds. 
From the panel, we can also see two clear regimes for the noise evolution (c.f. \citealt{GMK20}): (i) on small-scales, $k\gsim 0.1$ Mpc$^{-1}$, the noise is dominated by thermal noise, and increases strongly with redshift independently of the cosmic signal.  On large-scales, $k\lsim 0.1$ Mpc$^{-1}$, although the noise is generally still dominated by the (smoothly-evolving) thermal noise, the 
cosmic variance begins to have a non-negligible contribution.  As a result, the noise structure can be seen to trace the structure in the cosmic signal from the top right panel on large scales.

In the bottom right panel, we show the corresponding S/N (i.e. the ratio of the top right and bottom left panels).  From this, we clearly see that the highest S/N of order $\sim$ 10 occurs during the three large-scale peaks in the signal, corresponding to the EoR, EoH and WF coupling epochs.

\subsection{Bias and scatter of the cosmic 21-cm power spectrum}

Keeping the same astrophysical parameters, we run smaller box simulations and quantify the impact of the missing large-scale modes.  In Table \ref{tab:sim_list} we list the box sizes of the various simulations, including the number of independent realizations (different initial seeds) performed.  We keep the same cell resolution, $L_{\rm cell} \approx 1.5$ Mpc, for all simulations in this convergence study.

\begin{figure*}
\includegraphics[width=0.45\textwidth, height = 6 cm]{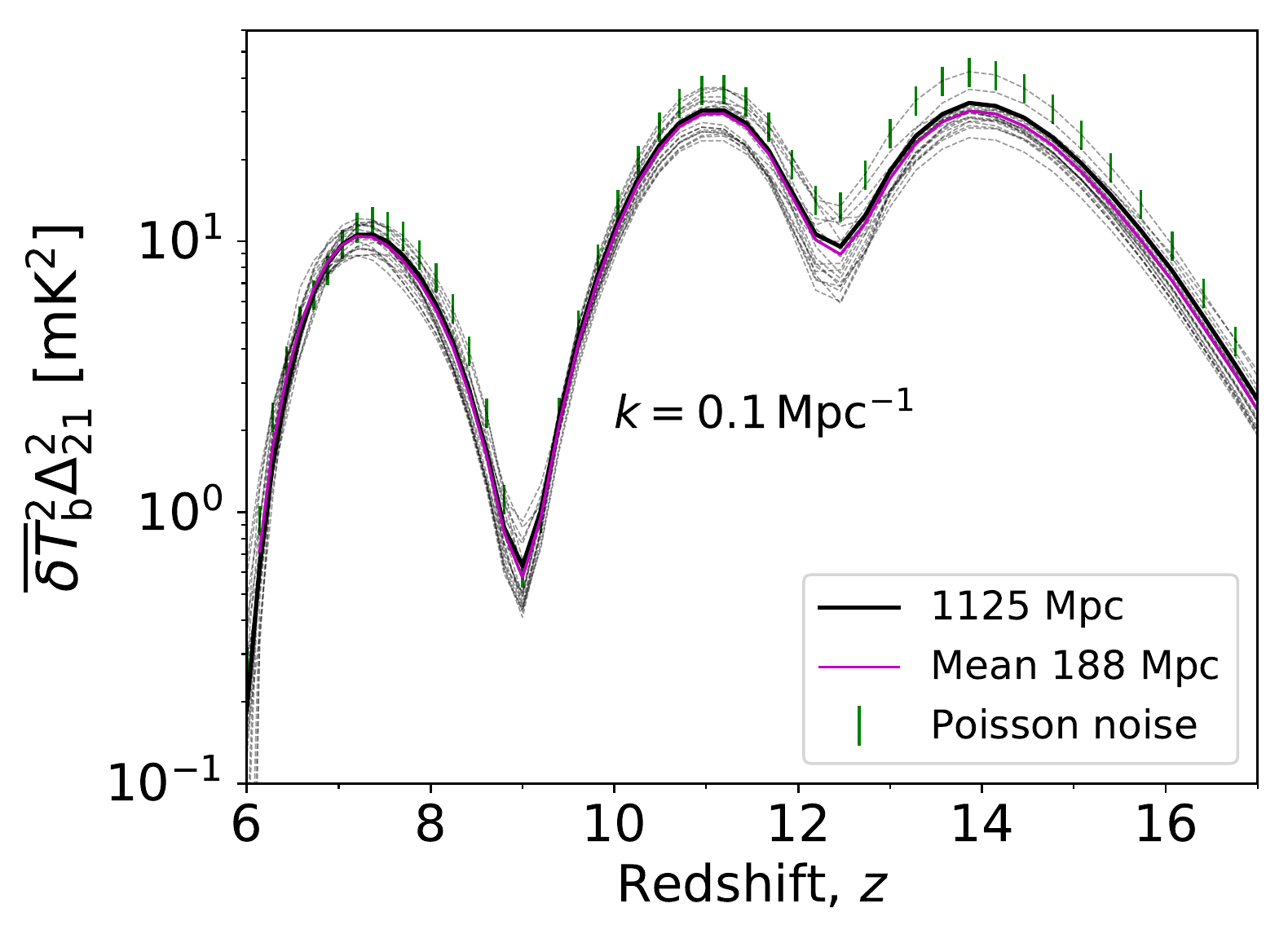}.
\includegraphics[width=0.45\textwidth, height= 6 cm]{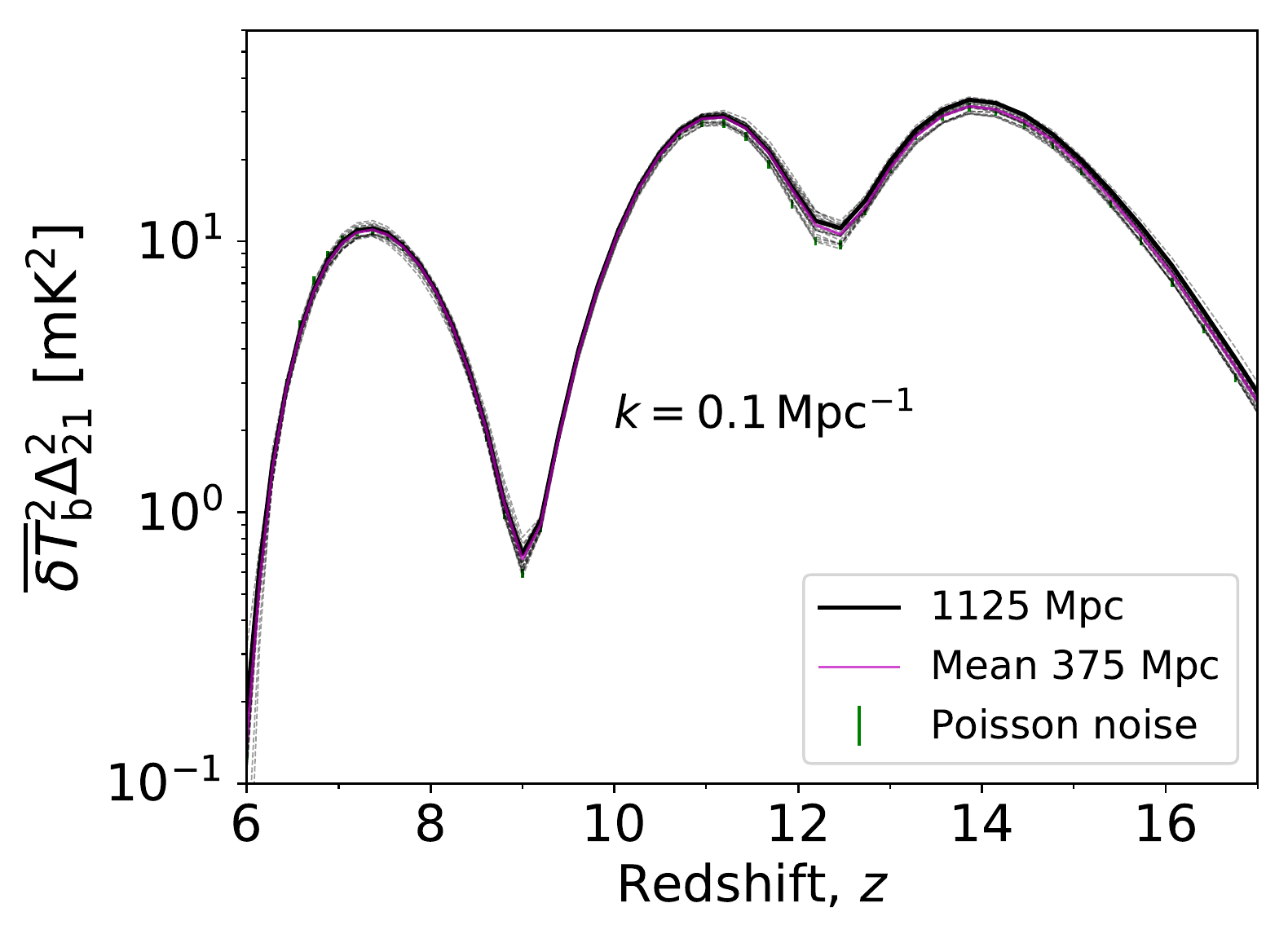}
\caption{ The redshift evolution of the cosmic 21-cm PS at $k=0.1$ Mpc$^{-1}$ for various realizations, using the same cosmological and astrophysical parameters.  Gray dashed curves correspond to realizations of $L_{\rm box}$ = 188 Mpc ({\it left panel}) and 375 Mpc ({\it right panel}), while the corresponding mean over all of the realizations is shown as a solid purple curve.  The reference simulation of $L$ = 1125 Mpc is shown with a black solid curve in each panel.  The green error bars on one of the simulations in each panel correspond to the Poisson uncertainty on the PS from a single small-box realization (from averaging over a discrete number of $k$-modes in Fourier space).
  The figure illustrates two expected trends: (i) a scatter in the measured 21-cm power from different realizations, which decreases with increasing box size; and (ii) a bias from the missing large-scale modes, with most of the small-box simulations having smaller large-scale power during the Cosmic Dawn.
  }
\label{fig:PSevolution}
\end{figure*}
The box sizes used are also illustrated in the top left panel of Fig. \ref{fig:ref}.  Even by eye, one can see notable structure in the 21-cm maps on scales comparable to the smallest box sizes.  We point out that even our smallest box size, $L \sim 200$ Mpc, encloses a volume which is a factor of $\gsim10$ larger is currently accessible with state-of-the-art numerical simulations resolving atomically-cooled galaxies (e.g. \citealt{Dixon16, Ocvirk18, DTC19}; see also the review in \citealt{TG11}).

We further illustrate the impact of limited box sizes in Fig. \ref{fig:PSevolution}.  Here the gray curves show the redshift evolution of the large-scale power for all of the realizations of our box sizes: 188 Mpc in the left panel and 375 Mpc in the right panel.  The solid purple curves show the mean over all of the realizations.  The reference large-scale simulation is shown with a black curve in both panels.  The Poisson uncertainty on the PS (i.e. the uncertainty on the mean amplitude from sampling a limited number of modes in Fourier space around $|{\bf k}|\approx0.1$ Mpc$^{-1}$), is denoted with green error bars for one of the realizations in each panel.  We note that the box-to-box scatter is larger than this Poisson uncertainty even for a such relatively large-scale mode of $k\approx0.1$ Mpc$^{-1}$ (corresponding to a comoving length of $\lambda = 2 \pi/k \approx$ 60 Mpc).

Figure \ref{fig:PSevolution} confirms two expected trends (e.g. \citealt{BL04, Iliev2014}): (i) the variance of the PS from different realizations decreases with increasing box size; and (ii) the PS constructed from smaller boxes {\it on average} underestimates the amount of 21-cm structure (i.e. the purple curves are lower than the black curves).


We quantify the bias of (ii) for all of our simulations by computing the fractional difference in power between the reference simulation and the smaller box simulations:
\begin{equation}
\label{eq:bias}
\langle\delta P(L, k, z)\rangle \equiv \biggl< \frac{P_{\rm L, i} - P_{\rm ref}}{P_{\rm ref}} \biggr>_{N_{\rm real}} ~ .
\end{equation}
Here,  $P_{\rm ref} (k, z)$ corresponds to the PS of our reference, 1125 Mpc large-scale simulation, $P_{L, i} (k, z)$ to the PS of a given realization $i$ with box length $L$, and the averaging is performed over all $N_{\rm real}$ realizations of that box size.  In the fourth column of Table \ref{tab:sim_list} we list $\langle \delta P (L) \rangle$ evaluated at $z=14$ when the $k=0.1$ Mpc$^{-1}$ power peaks, corresponding to the Cosmic Dawn epoch when fluctuations in the Ly$\alpha$ coupling dominate the signal.  Box sizes lower than $L\lsim300$ underestimate the power at peak CD signal by up to $\sim$7--9\%.  This bias decreases to $\sim$ 1 \% for the 563 Mpc simulations.  We confirm that this error is also much smaller during the EoR, dropping to $\sim$ 1\% even for $L\lsim 300$ Mpc boxes.  This is understandable since ionization fluctuations during the EoR occur on smaller scales than the temperature and Ly$\alpha$ coupling fluctuations during the CD.


\begin{table}[!h]
        \centering
        \caption{List of the smaller box simulations used in this work.  Columns correspond to: (i) the side length of the simulation, $L$; (ii) the number of cells, $N_{\rm cell}$; (iii) the number of independent realizations, $N_{\rm real}$; (iv) the fractional bias in the 21-cm PS (c.f. eq. \ref{eq:bias}), averaged over all realizations, computed at the peak of the large-scale power, $(k=0.1 {\rm Mpc}^{-1}, z=14)$; and (v) the mean of the S/N-weighted error, in units of the total noise (c.f. eq. \ref{eq:SN_weighting}; note that the median of this error is denoted with horizontal lines in Fig.\ \ref{fig:SN_weighted}).}
        \label{tab:sim_list}
        \begin{tabular}{rrrcc} 
                \hline
                $L (\rm Mpc)$ &
                $N_{\rm cell}$ &
                $N_{\rm real}$ &
                $\langle \delta P \rangle$ (\%) &
                $\Delta P_{\rm S/N, i}$  $(\sigma_{\rm tot})$\\
                \hline
                187.50 & 128$^3$ &  20 & -6.6 $\pm 2.7 (1\sigma)$ & 0.90 $\pm 0.16 (1\sigma)$ \\
                281.25 & 192$^3$ &  20 & -8.8 $\pm 1.6 (1\sigma)$ & 0.70 $\pm 0.14 (1\sigma)$\\
                375.00 & 256$^3$ & 20 & -5.6 $\pm 1.2 (1\sigma)$  & 0.48 $\pm 0.16 (1\sigma)$\\
                468.75 & 320$^3$ & 10 & -2.5 $\pm 1.2 (1\sigma)$ & 0.43 $\pm 0.08 (1\sigma)$\\
                562.50 & 384$^3$ & 10 & -1.4 $\pm 1.1 (1\sigma)$ & 0.30 $\pm 0.07 (1\sigma)$\\
                \hline
        \end{tabular}
\end{table}

\subsection{Convergence in the signal}

\begin{figure*}
\includegraphics[width=0.33\textwidth]{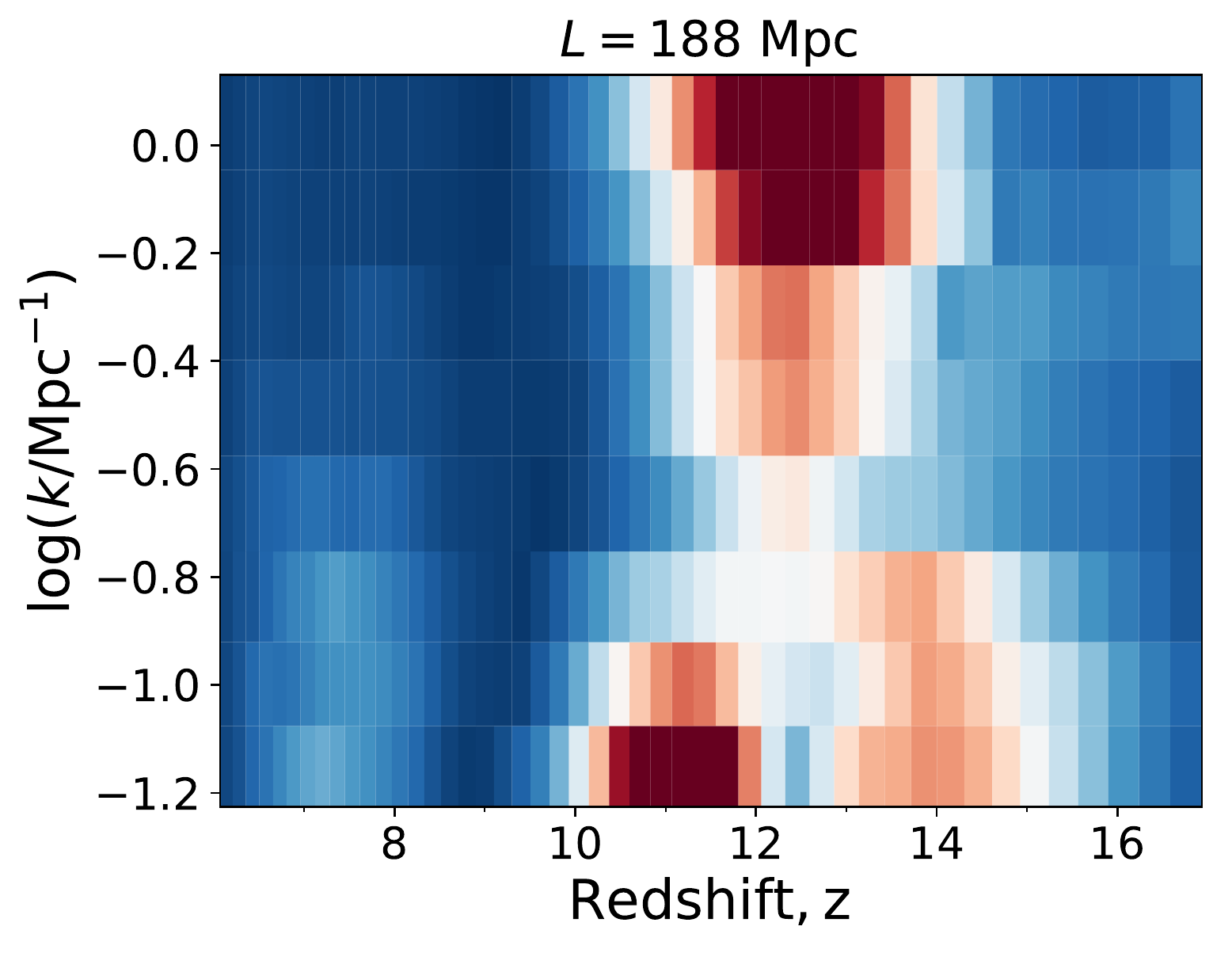}
\includegraphics[width=0.33\textwidth]{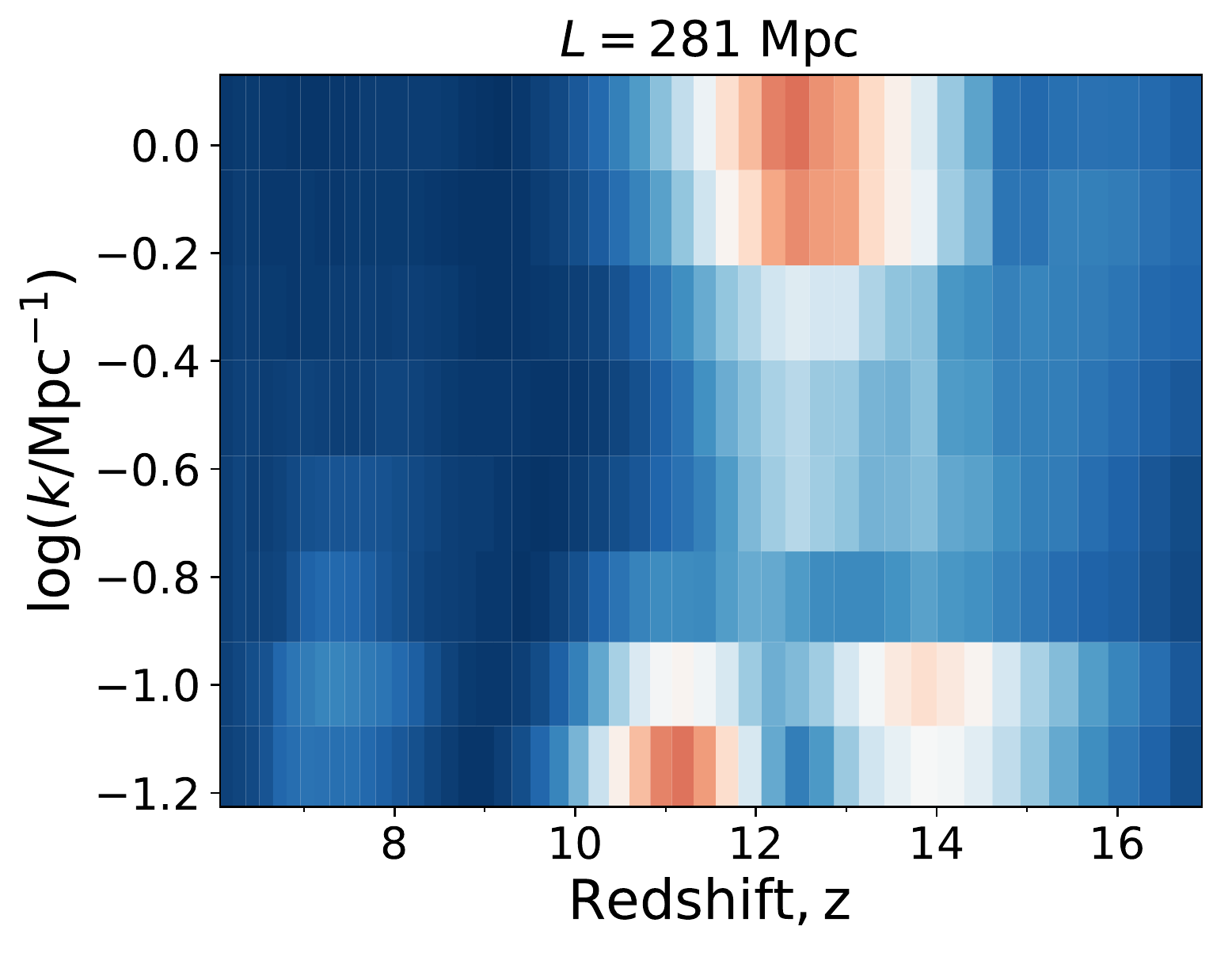}
\raisebox{3ex}{\includegraphics[width=0.325\textwidth]{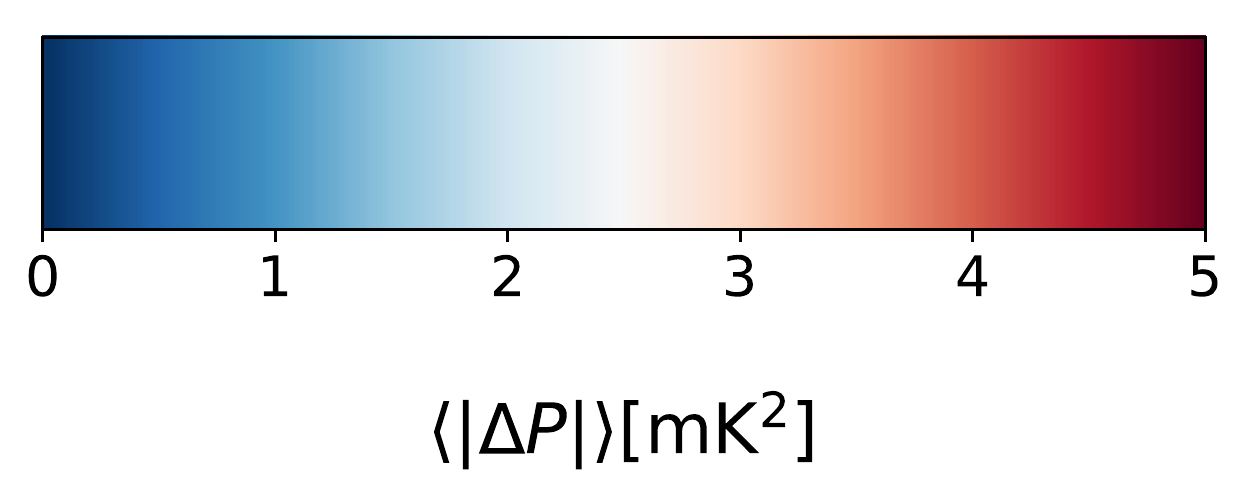}}
\includegraphics[width=0.33\textwidth]{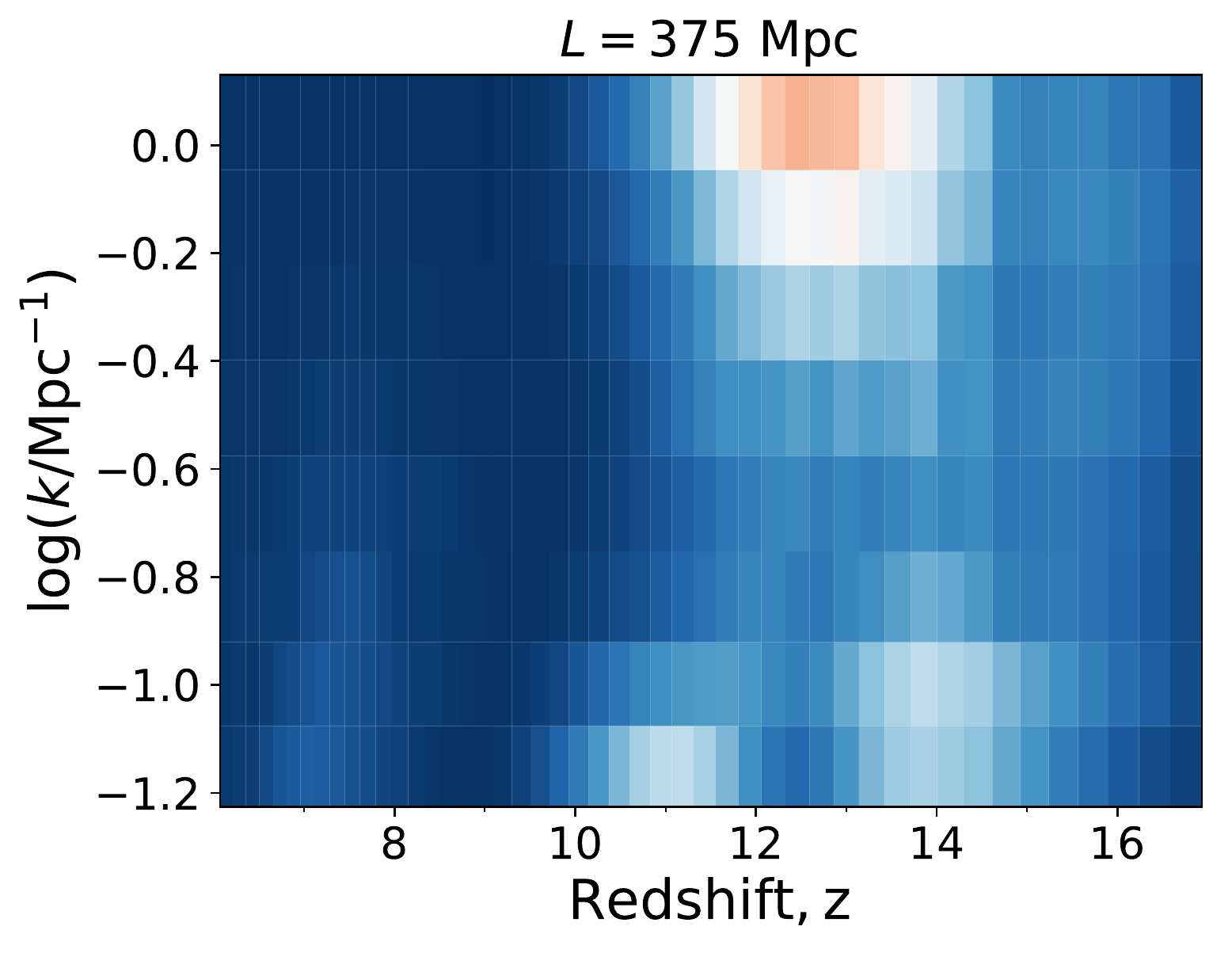}
\includegraphics[width=0.33\textwidth]{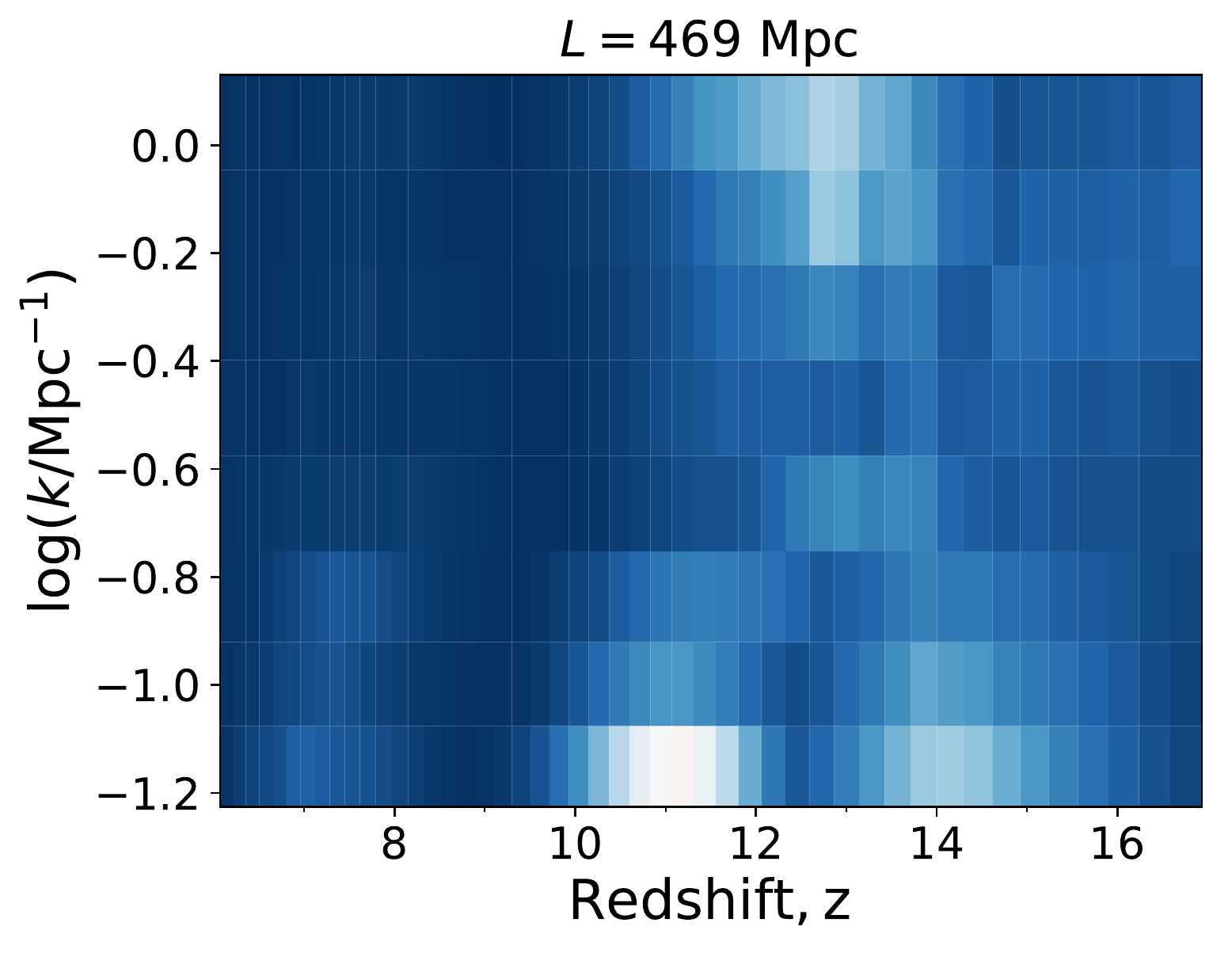}
\includegraphics[width=0.33\textwidth]{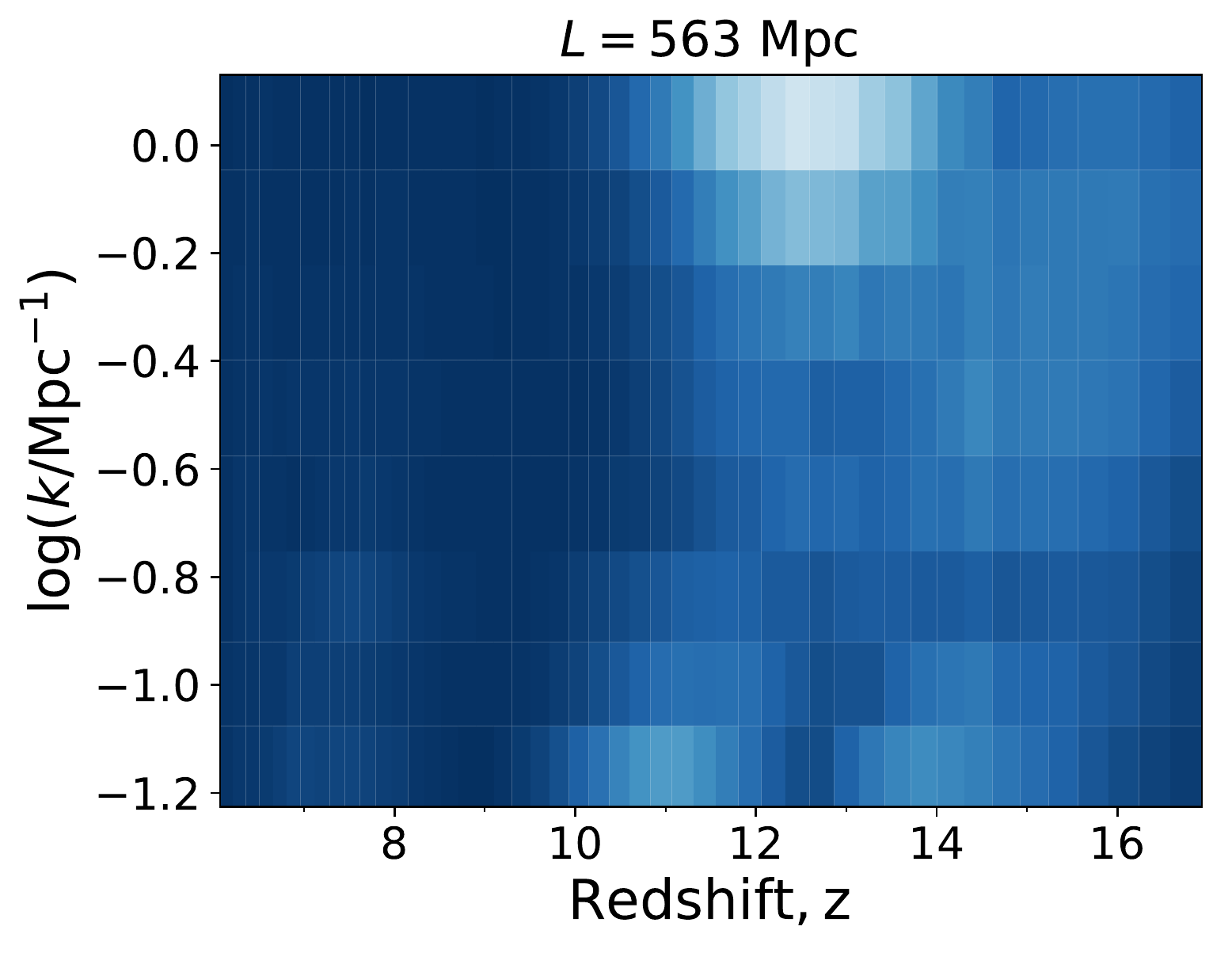} 
\caption{Absolute difference in the power spectra amplitudes of $L=$ 188, 281, 375, 469, and 562 Mpc boxes with respect to the reference 1125 Mpc box, averaged over different realizations of the initial conditions.  The average difference for our $L=188$ Mpc simulation goes up to $\sim 7$ mK$^2$.}
\label{fig:abs_diff}
\end{figure*}

In order to quantify the convergence of the small box simulations with respect to the reference, we calculate the average of absolute differences:
\begin{equation}
\label{eq:absdif}
\langle |\Delta P(L,k,z)| \rangle \equiv \langle  | P_{\rm L, i} - P_{\rm ref} | \rangle_{N_{\rm re    al}} ~ .
\end{equation}
$\langle|\Delta P(L, k, z)| \rangle$ is a measure of the scatter in the PS amplitude at $(k, z)$, for a simulation of box length, $L$.
We plot these absolute PS differences in Fig. \ref{fig:abs_diff}.

On large scales, the most significant differences in the PS occur during the three astrophysical epochs: (i) EoR at $z\sim7$; (ii) EoH at $z\sim11$; (iii) WF coupling at $z\sim14$.  These correspond to the three peaks of the large-scale power (c.f. Fig. \ref{fig:ref}).  Of these, the EoH has the largest scatter. On small-scales, the largest scatter occurs during the cosmic dawn, again tracing the amplitude of the power spectrum.

The PS differences in the smallest box simulations, $L=188$ Mpc reach values of $\langle |\Delta P| \rangle \sim$ 7 mK$^2$.  As expected, there is a clear decrease in the scatter with increasing simulation box size.

However, not all scales and redshifts are equally relevant from an observational point of view. For example, on small scales or at high redshifts, thermal noise can be quite high, making the 21-cm signal unobservable even with SKA1-low (c.f. Fig. \ref{fig:ref}).  Since these modes are unobservable, they should be less important when estimating convergence criteria.
\begin{figure*}
\includegraphics[width=0.33\textwidth, height=5cm]{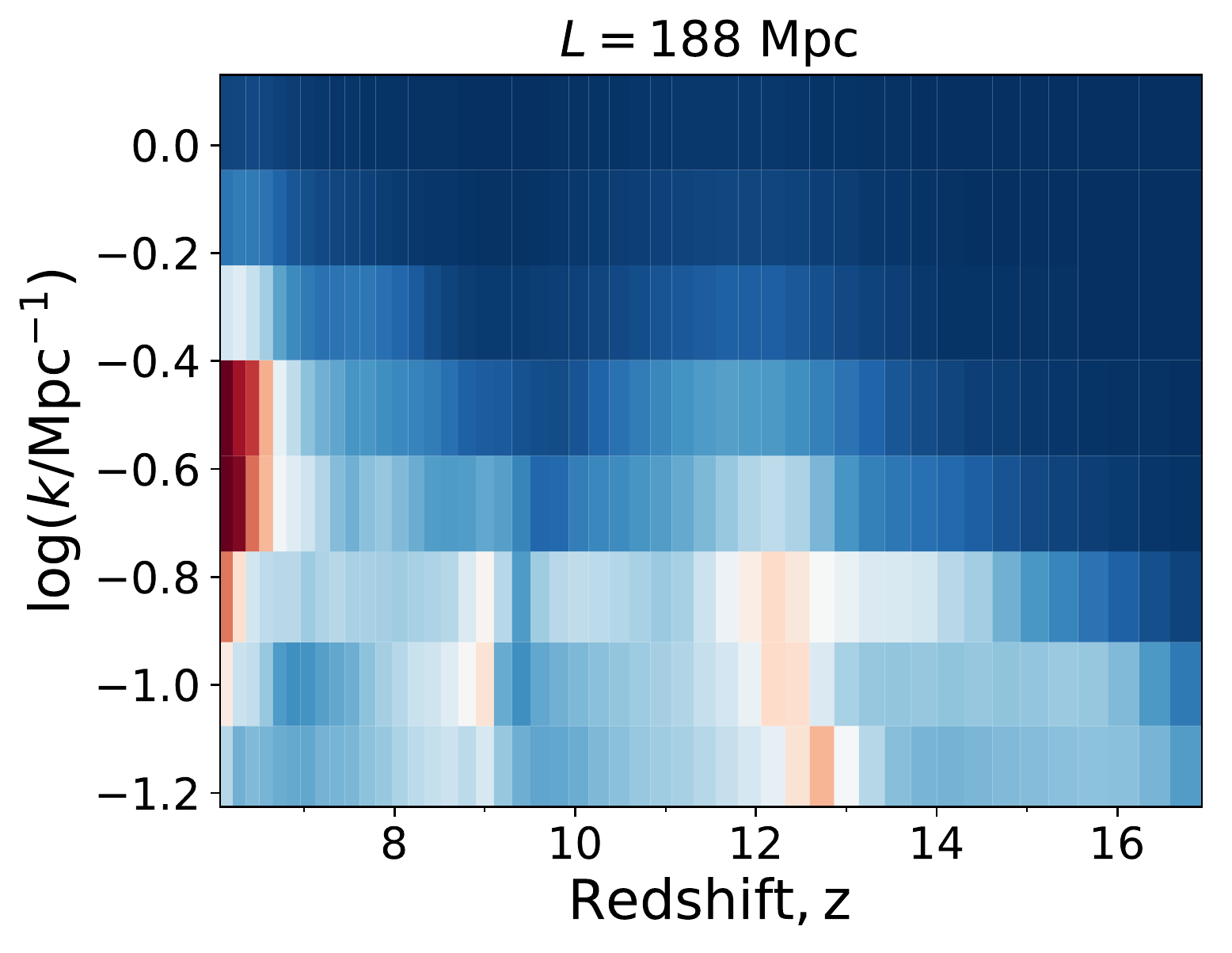}
\includegraphics[width=0.33\textwidth, height=5cm]{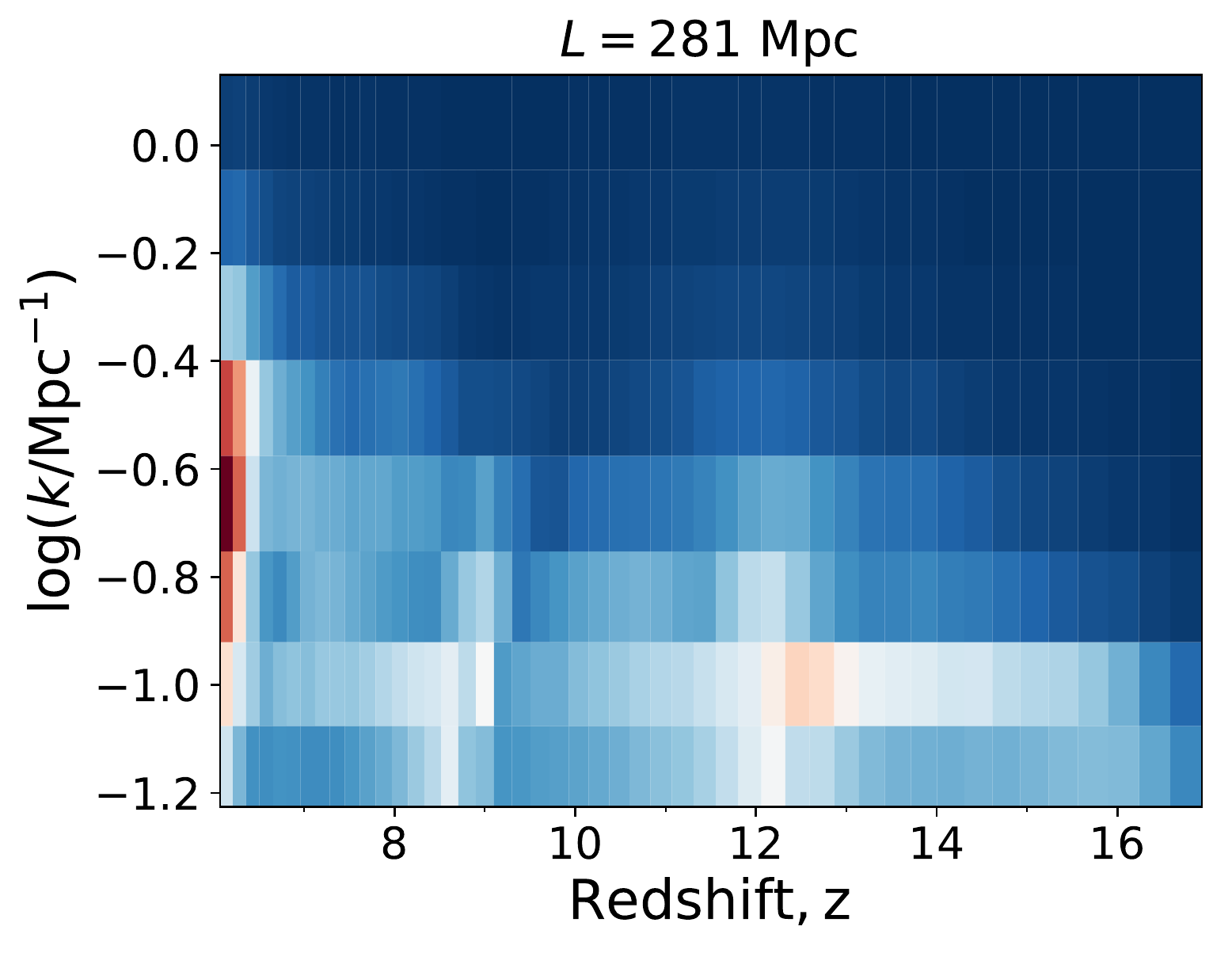}
\raisebox{3ex}{\includegraphics[width=0.325\textwidth]{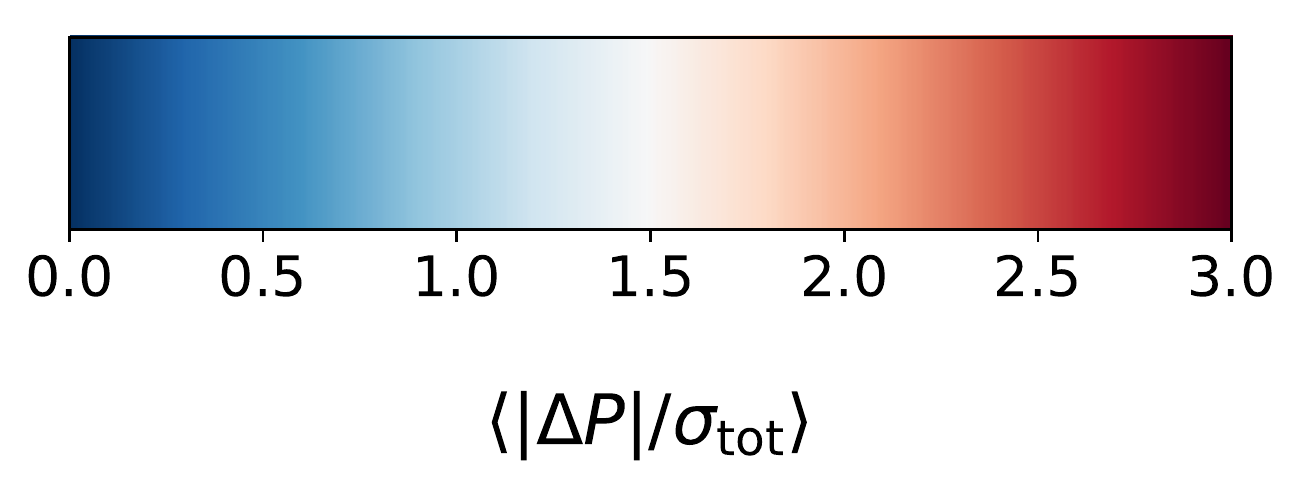}}
\includegraphics[width=0.33\textwidth, height=5cm]{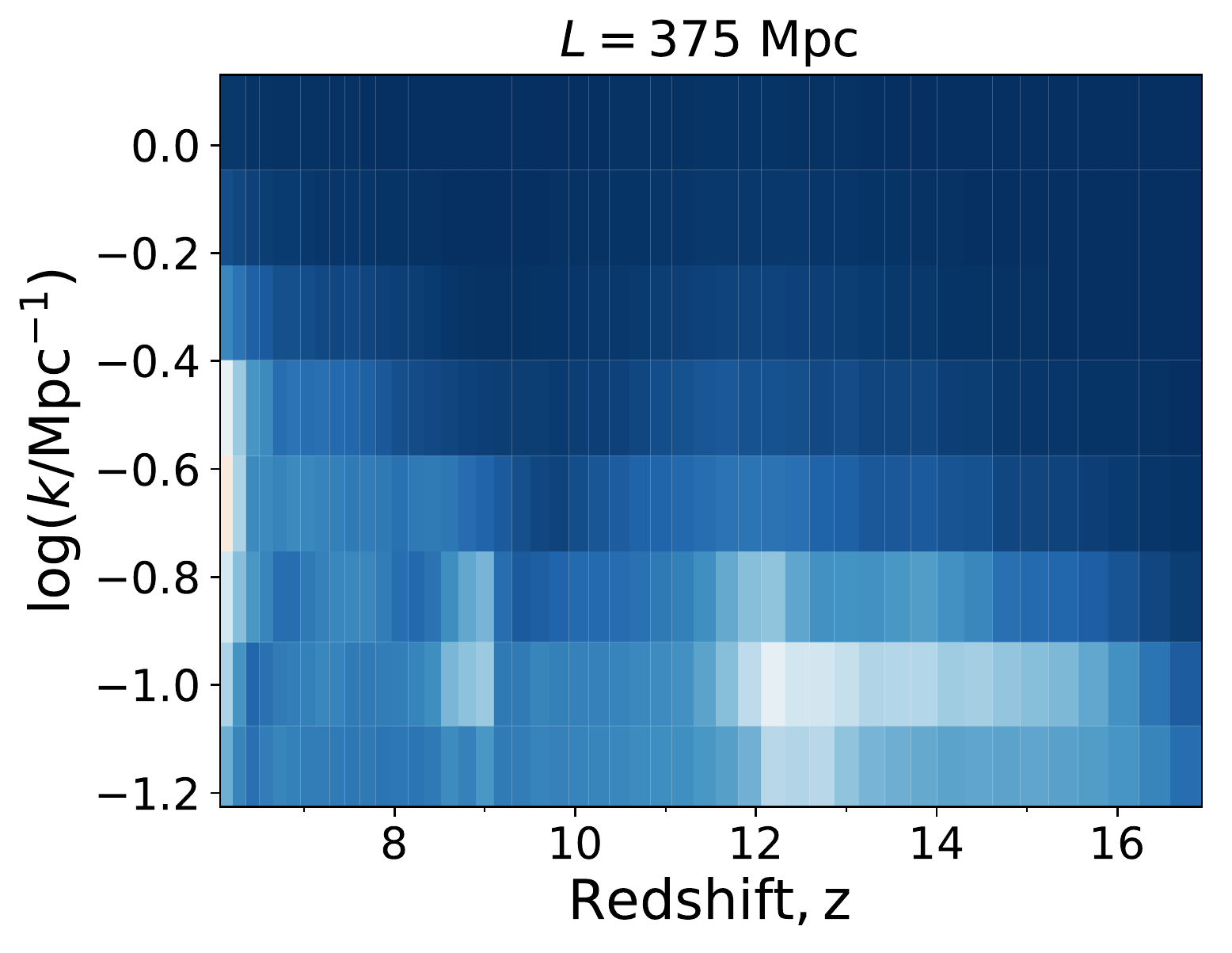}
\includegraphics[width=0.33\textwidth,height=5cm]{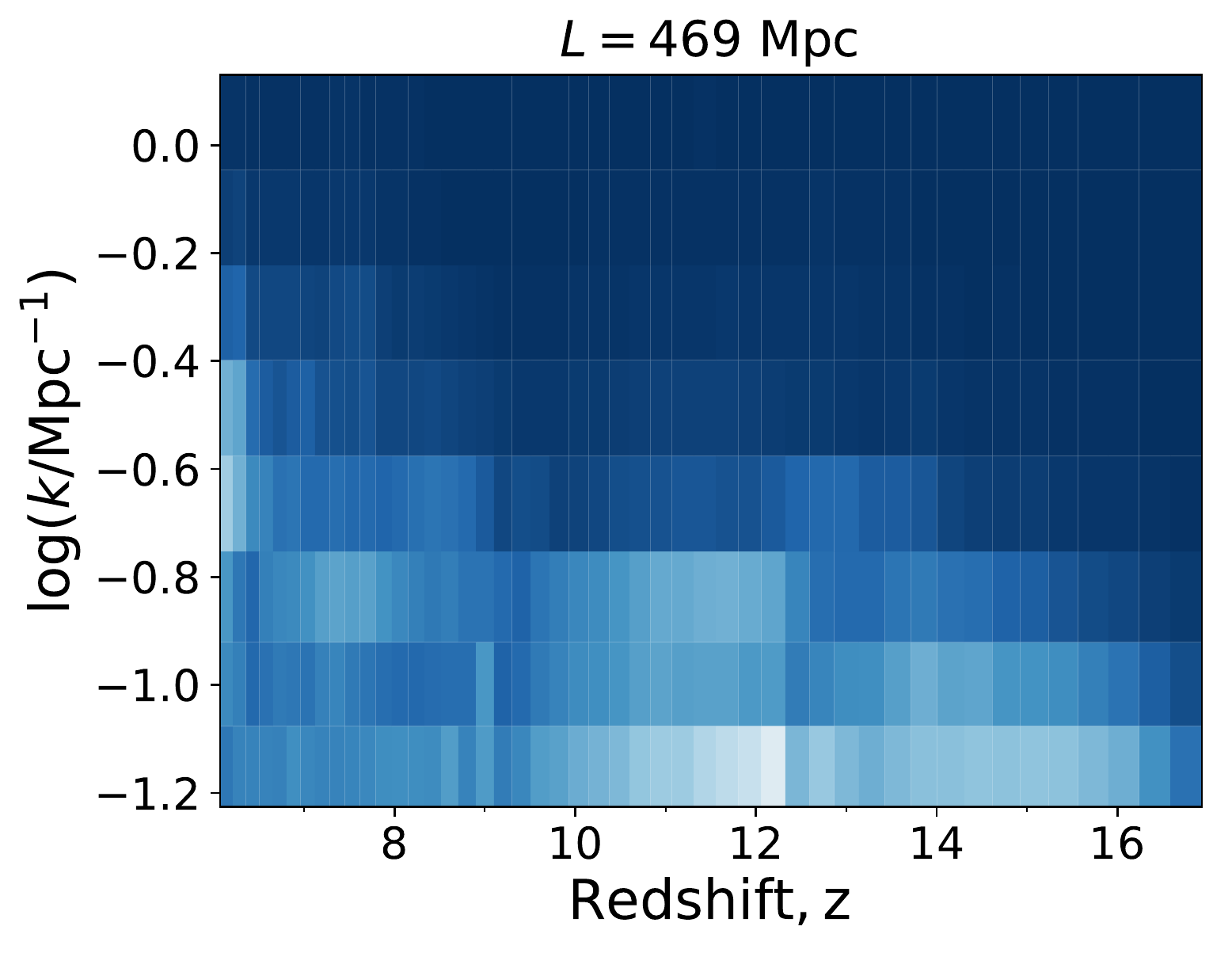}
\includegraphics[width=0.33\textwidth,height=5cm]{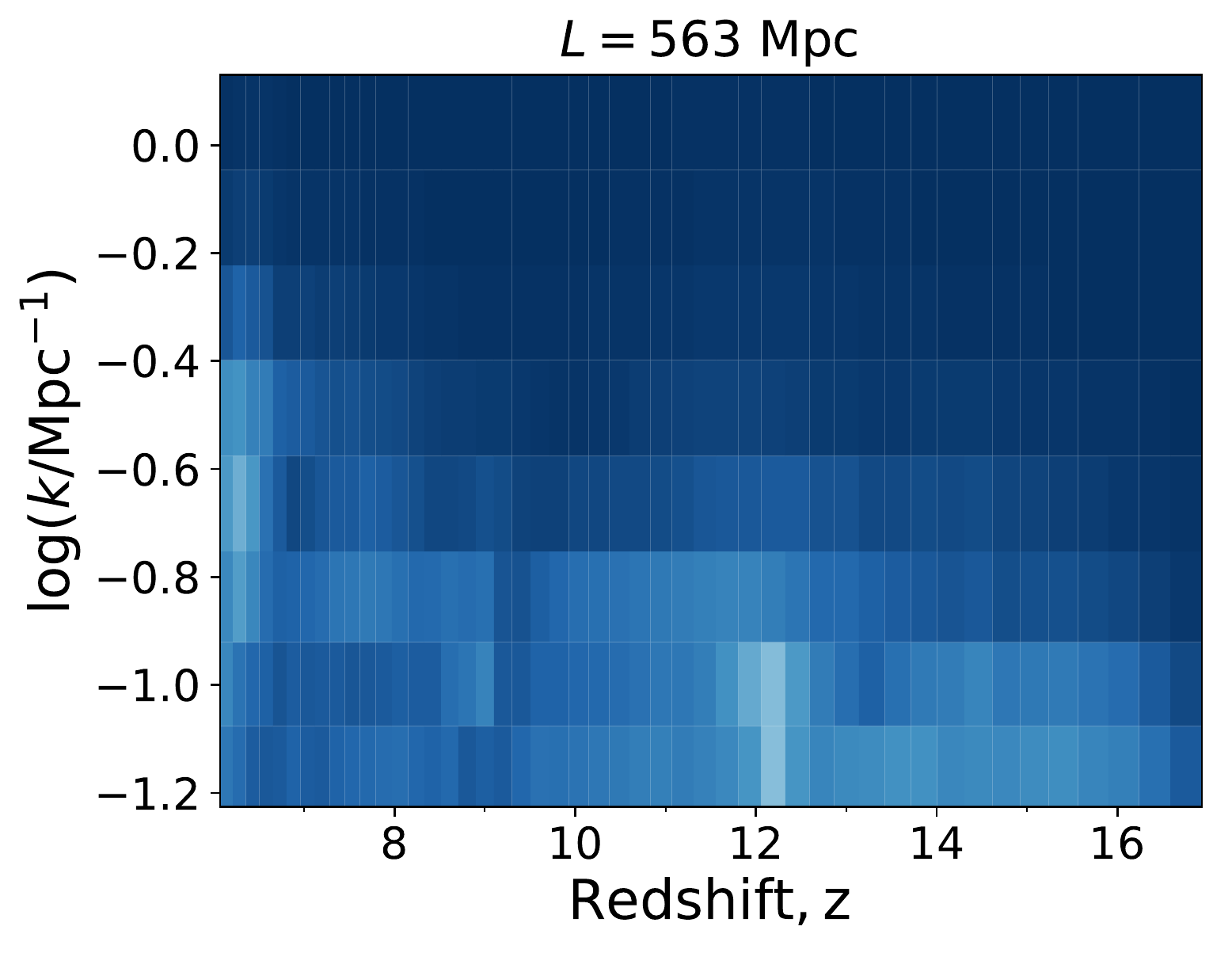}
\caption{Same as Fig. \ref{fig:abs_diff}, but dividing by the total noise,\ $\sigma_{\rm tot} $ in each $(k, z)$ bin.  The total noise, $\sigma_{\rm tot} $, includes the thermal and sample variance of the reference simulation from eq. (\ref{eq:noise}) with the Poisson sample variance of the small box realizations added in quadrature.  The average error in the $L=188$ Mpc simulations goes up to $\sim 4 \sigma_{\rm tot}$.}
\label{fig:noise_diff}
\end{figure*}

With this in mind, in Figure \ref{fig:noise_diff} we re-plot the average PS differences, but in units of the total r.m.s. noise: $\langle |\Delta P| / \sigma_{\rm tot} \rangle$.  Here $\sigma_{\rm tot}$ corresponds to the total noise, including thermal and sample variance of the reference simulation from eq. (\ref{eq:noise}) with the Poisson sample variance of the small box realizations added in quadrature.

From Fig. \ref{fig:noise_diff} we see that the scatter in the PS differences expressed in terms of the total noise is largest on large-scales where the thermal noise is the smallest.
In particular, the late stages of the EoR and the CD show differences of up to $\sim 4 \sigma_{\rm tot}$ for simulations of box sizes $L\sim$ 200 - 300 Mpc.  These differences decrease to below 1 $\sigma_{\rm tot}$ for our largest box sizes.


\begin{figure*}
\includegraphics[width=0.7\textwidth]{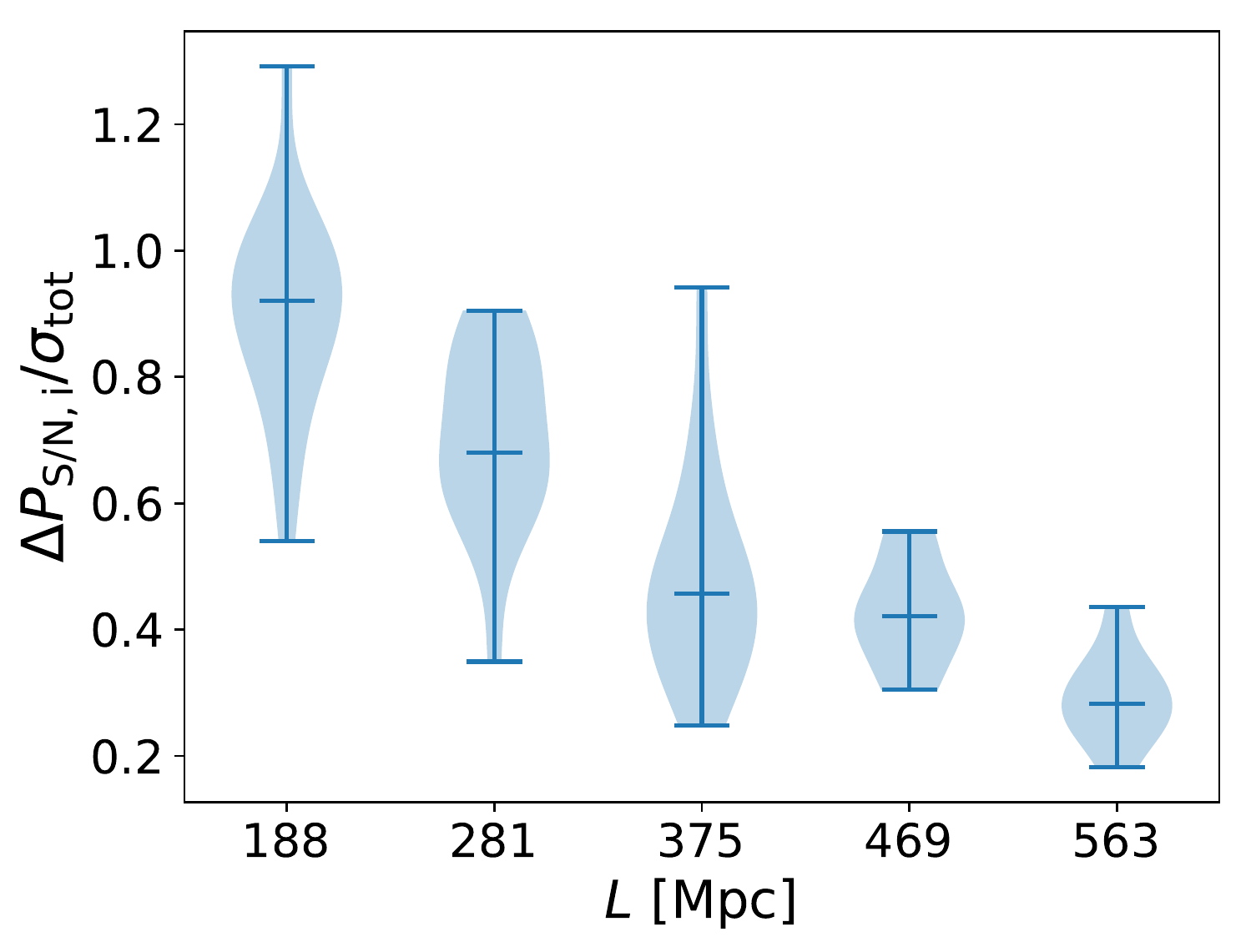}
\caption{Violin plots of the S/N weighted average over $(k, z)$ of the absolute difference in PS amplitude, in units of the total noise (see equation \ref{eq:SN_weighting}).  The middle horizontal lines denote the median of the distributions over realizations, $i$, while the bars enclose the full extent.
For a box size of 188 Mpc, the median, S/N-weighted PS error is $\sim0.9 \sigma_{\rm tot}$, while the r.m.s. (1$\sigma$) of the distribution is 0.16$\sigma_{\rm tot}$.
Both the median and the spread of the S/N-weighted PS error decrease with increasing box size.}
\label{fig:SN_weighted}
\end{figure*}

Finally, we marginalize the PS differences over $(k, z)$,  weighing by the S/N.  Specifically, we compute:
\begin{equation}
\Delta P_{\rm S/N, i} =
\frac{ \int_z\int_k [{\rm S/N}] \frac{ | P_{\rm L, i} - P_{\rm ref} |}{\sigma_{\rm tot}}  dk ~ dz}{\int_z\int_k [{\rm S/N}] ~ dk ~dz} ~ .
\label{eq:SN_weighting}
\end{equation}
Here S/N$= P_{\rm ref} / \sigma_{\rm ref}$ refers to the reference mock observation, shown in the bottom right panel of Fig. \ref{fig:ref}.

Equation (\ref{eq:SN_weighting}) provides a single number for a given small-box realization, $i$, corresponding to the S/N weighted average over $(k, z)$ of the absolute difference in PS amplitude, in units of the total noise.  We plot the distributions of $\Delta P_{\rm S/N, i}$ in Figure \ref{fig:SN_weighted} for all box sizes.  The distributions over realizations, $i$, are shown with violin plots.

As expected, both the median and the spread of this S/N-weighted PS error decreases with increasing box size.
For a box size of 188 Mpc, the median, S/N-weighted PS error is $0.9 \sigma_{\rm tot}$, while the r.m.s. (1$\sigma$) of the distribution is 0.16$\sigma_{\rm tot}$.
For larger box sizes, none of our realizations have a S/N-weighted PS error greater than $1 \sigma_{\rm tot}$.

\section{Conclusions}
\label{sec:conc}

Interferometric observations of the cosmic 21-cm signal are set to revolutionize
our understanding of the Epoch of Reionization (EoR) and the Cosmic Dawn (CD).  However, interpreting these observations relies on our ability to accurately model the large-scale cosmological signal.

The first galaxies are likely very rare and biased, with their abundances modulated by long-wavelength modes of the density field (e.g. \citealt{BM96_algo}).  Moreover, the radiation fields from these galaxies interact with the IGM over a large range of scales (e.g. \citealt{Pritchard2007}).  Therefore, the limited volume of 21-cm simulations can underestimate the amount of structure in the cosmic 21-cm signal (e.g. \citealt{BL04, Iliev2014}).

In this work we quantify the minimum box size for simulating the power spectrum of the cosmic 21-cm signal.  Using the public code $\cmfast$, we perform multiple realizations of the cosmic 21-cm signal for a range of box sizes.  We quantify convergence with respect to a mock observation of box length $1125$ Mpc, with thermal noise computed for a 1000h observation with SKA1-low assuming the optimistic foreground scenario of \citet{Pober2014}.

We find that simulations of box lengths $ L \lsim 200$ Mpc typically do not show a bias in the PS during the EoR; however they do tend to underestimate the large-scale power during the earlier epoch of CD by $\sim$ 7\%. There is also notable scatter between different realizations.  As expected, both the bias and scatter decreases with increasing box size. 

We quantify the absolute difference in the error between the power spectra from small-box realizations and the reference simulation.  This error, averaged over multiple realizations, reaches values of up to $\sim$ 7 mK$^2$ for the $L=188$ Mpc simulation. We also compute this error in terms of the total noise, accounting for the fact that some modes are easier to detect than others.  The error reaches values of $\sim 4 \sigma_{\rm tot}$ for the smallest simulations.

Finally, we marginalize the error over all $(k, z)$ modes, weighted by the corresponding S/N.  We conclude that box lengths of $L \gsim$ 250 Mpc are needed to converge at the level of $\lsim$ 1 $\sigma$ of the total noise.  This corresponds to simulation volumes $\gsim$ 10 times larger than state-of-the-art radiative transfer simulations that resolve atomic cooling galaxies.

\section{Acknowledgements}
We thank Yuxiang Qin and Jaehong Park for helpful discussions.
This work was supported by the European Research Council (ERC) under the European Union's Horizon 2020 research and innovation programme (grant agreement No. 638809 - AIDA - PI: Mesinger). The results presented here
reflect the authors' views; the ERC is not responsible for their use.




\bibliographystyle{mnras}
\bibliography{ms} 




\appendix

\section{Dependence on astrophysics}
\label{Sec:Extreme_model}


\begin{figure}
\includegraphics[width=0.45\textwidth]{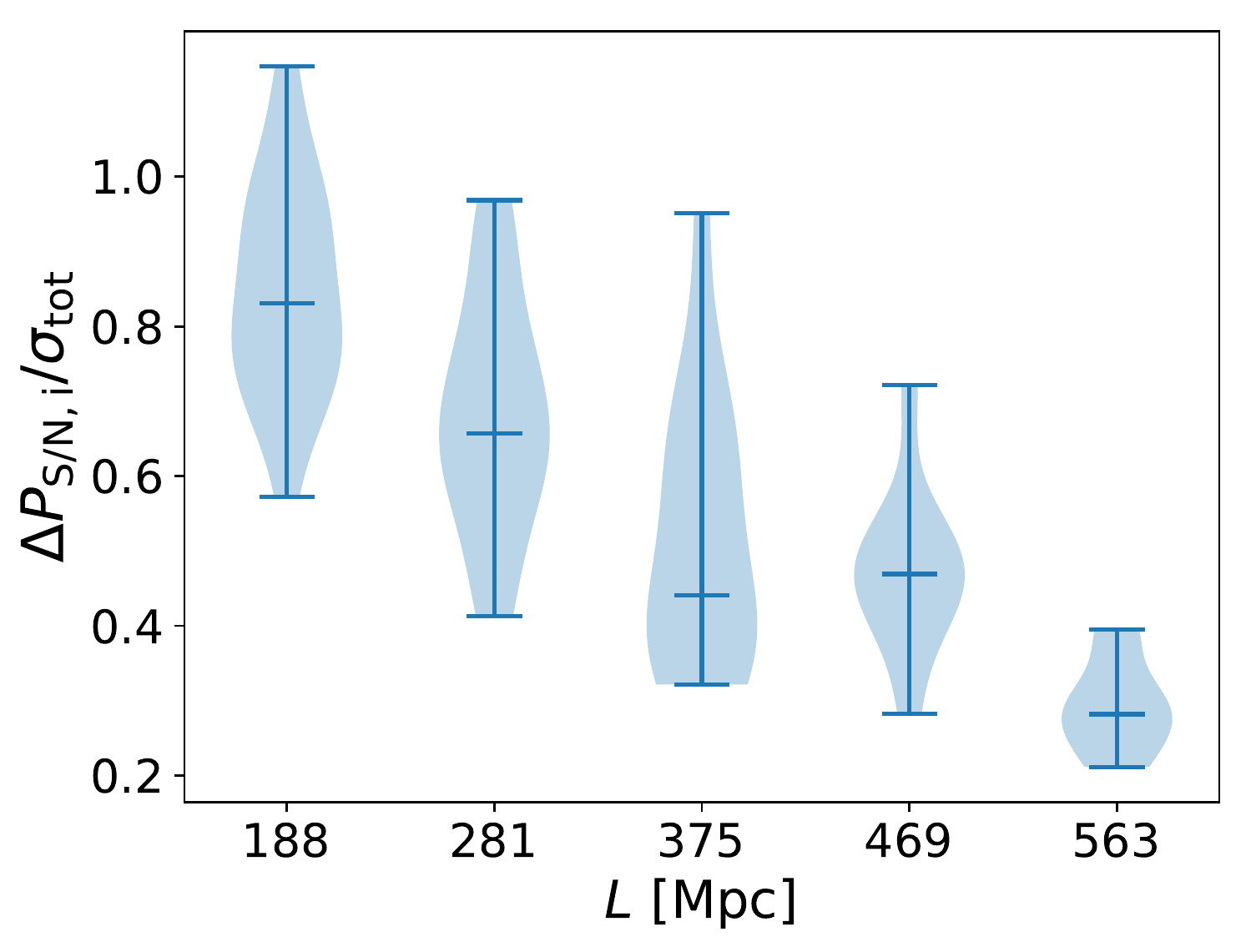}
\caption{Analogous to Fig. \ref{fig:SN_weighted}, but computed for simulations with $M_{\rm turn}= 5\times10^9 M_\odot$, which is a factor of ten larger than our fiducial choice.  Although the variance in the small box simulations is slightly larger, we recover similar trends.}
\label{fig:SN_extreme}
\end{figure}

In order to test the dependence of our conclusions on the astrophysical parameters, we perform another convergence test, but increasing $M_{\rm turn}$ by a factor of 10.  In other works, we take $M_{\rm turn} = 5 \times 10^9  M_{\odot}$, keeping the other parameters the same. Increasing $M_{\rm turn}$ corresponds to increasing the bias of star-forming galaxies, delaying all astrophysical epochs and increasing the PS amplitude (e.g. \citealt{GM17_21CMMC}).

We perform $N_{\rm real}=$ 20, 20, 15, 10, 10 realizations of $L = $ 188, 281, 375, 469, 563 Mpc simulations (respectively),
comparing them to a 1125 Mpc mock observation generated with the same astrophysics.   We plot the main results in Fig. \ref{fig:SN_extreme}, which is analogous to Fig. \ref{fig:SN_weighted} for our fiducial astrophysics.

We see that the variance in the smallest box sizes has increased for this model.  Overall, the trends are roughly the same as those noted for the fiducial model.  Specifically, we again conclude that box sizes of $\gsim$ 250 Mpc are required for the PS to converge to within 1 $\sigma$ of the total noise.  This is mostly because although the large-scale 21-cm PS of the $M_{\rm turn} = 5 \times 10^9  M_{\odot}$ model is larger, so is the corresponding sample variance component of the noise.  Therefore the convergence criteria expressed in terms of the total noise is comparable.


\bsp	
\label{lastpage}
\end{document}